\documentclass[prd,aps,floats,floatfix,superscriptaddress,preprintnumbers,
showpacs,eqsecnum,nofootinbib,twocolumn]{revtex4}
\usepackage{latexsym,array,theorem,mathrsfs,subfigure,bm,float}

\usepackage{psfrag}
\usepackage{amsfonts,amsmath,amssymb,latexsym,array,afterpage,
theorem,mathrsfs,bm,float,epsfig,color,graphicx,tabularx,here,multirow}

\newcommand{\nn}{\nonumber \\}
\newcommand{\bea}{\begin{eqnarray}}
\newcommand{\ena}{\end{eqnarray}}
\newcommand{\beann}{\begin{eqnarray*}}
\newcommand{\enann}{\end{eqnarray*}}

\newcommand{\lsim}{\, \mbox{\raisebox{-1.ex}
{$\stackrel{\textstyle<}{\textstyle\sim}$}}\,}

\newcommand{\nablag}{\overset{ \scriptsize (g)}{\nabla}}

\newcommand{\RB}{\bar{\mathstrut R}}

\newcommand{\gB}{\bar{ g}}
\newcommand{\fB}{\bar{\mathstrut f}}
\newcommand{\BoxB}{\bar{\mathstrut \Box }}
\newcommand{\nablaB}{\bar{\mathstrut \nabla}}

\begin{document}

\baselineskip=12pt

\preprint{IPMU 15-0082}
\title{Stability of the Early Universe in Bigravity Theory}
\author{Katsuki \sc{Aoki}}
\email{katsuki-a12@gravity.phys.waseda.ac.jp}
\affiliation{
Department of Physics, Waseda University,
Shinjuku, Tokyo 169-8555, Japan
}

\author{Kei-ichi \sc{Maeda}}
\email{maeda@waseda.ac.jp}
\affiliation{
Department of Physics, Waseda University,
Shinjuku, Tokyo 169-8555, Japan
}

\author{Ryo \sc{Namba}}
\email{ryo.namba@ipmu.jp}
\affiliation{Kavli Institute for the Physics and Mathematics of the Universe (WPI), 
The University of Tokyo Institutes for Advanced Study,
The University of Tokyo, Kashiwa, Chiba 277-8583, Japan
}

\date{\today}

\begin{abstract}
We study
the stability of
a spherically symmetric perturbation 
around the flat Friedmann-Lema$\hat{\i}$tre-Robertson-Walker spacetime
in the ghost-free bigravity theory, retaining nonlinearities of the 
helicity-$0$ mode of the massive graviton.
It has been known that, 
when the graviton mass is smaller than the Hubble parameter,
homogeneous and isotropic spacetimes suffer
from the Higuchi-type ghost or the gradient instability 
against the linear perturbation in the bigravity.
Hence, the bigravity theory has no healthy massless limit
for cosmological solutions
at linear level.
In this paper we show that the instabilities can be resolved by 
taking into account nonlinear effects of the 
scalar graviton mode
for 
an appropriate parameter space of
coupling constants.
The growth history in the bigravity 
can be restored to the result in general relativity
in the early stage of the Universe,
in which the St\"uckelberg fields are nonlinear
and there is neither ghost nor gradient instability.
Therefore, 
the bigravity theory has the healthy massless limit,
and cosmology based on it is viable even when
the graviton mass is smaller than the Hubble parameter.
\end{abstract}



\maketitle

\section{Introduction}

General relativity (GR) is now widely accepted as the low-energy effective theory of gravity and has passed a number of observational and experimental tests \cite{Will}. It can also be viewed as the unique theory of a massless spin-$2$ particle, namely graviton.
Although theories with a massive spin-2 field 
are one of the most natural extensions of GR,
such theories have suffered from many problems.
Fierz and Pauli proposed a massive spin-2 field theory,
which is known as the unique theory of linearized massive gravity free from a ghosty sixth degree of freedom \cite{FP}.
The Fierz-Pauli (FP) theory has appropriate five degrees of freedom, 
and there is no ghost mode on the Minkowski background.
However, the FP theory has no Newtonian limit even in a massless limit.
When we consider a naive massless limit in the FP theory,
the gravitational behaviour is not restored to the one in Newtonian gravity 
due to the existence of the extra gravitational degree of freedom,
which is called the van-Dam-Veltman-Zakharov (vDVZ) discontinuity \cite{vDVZ}.
Vainshtein then proposed that the vDVZ discontinuity can be  evaded by taking into account 
nonlinear mass terms,
and the extra mode is screened inside the so-called Vainshtein radius, to recover the standard gravitational interaction mediated only by the helicity-$2$ modes
\cite{original_Vainshtein}.
Boulware and Deser pointed out, however, that such nonlinear terms ruin the structure of the FP theory and reintroduce the ghost instability associated with the sixth degree of freedom \cite{BD}. Although this nonlinear ghost, often called Boulware-Deser (BD) ghost, appears in any simple nonlinear extensions of the FP massive gravity theory,
it was  shown in 2010 by de Rham et al. that the special choice of the mass term can eliminate such a ghost state at the decoupling limit \cite{dRGT}, and later the proof was extended to fully nonlinear orders \cite{gf1,gf2,gf3,gf4}.


However, this theory, often dubbed de Rham-Gabadadze-Tolley (dRGT) theory, has still been problematic in the context of cosmology.
It was revealed that the original dRGT theory does not admit any nontrivial flat or closed Friedmann-Lema\^{i}tre-Robertson-Walker (FLRW) universe,
if the fiducial metric for the St\"{u}ckelberg fields is Minkowski's one \cite{MassiveCosmo1}.
While this issue can be resolved either by open FLRW solutions  \cite{Gumrukcuoglu:2011ew} or by replacing the Minkowski fiducial metric with an FLRW one \cite{Gumrukcuoglu:2011zh}, it was later shown that all homogeneous and isotropic FLRW solutions in the dRGT theory are unstable due to either a linear ghost, called Higuchi ghost \cite{Higuchi}, or a new type of nonlinear ghost \cite{MassiveCosmo6}.
The ghost of the former type was found present already in the FP theory when it is constructed on a de Sitter background. The scalar graviton mode in the five degrees of freedom in the FP theory becomes a ghost when the graviton mass is below the so-called Higuchi bound, which is defined by a cosmological constant, leading to the conclusion that the theory has no healthy massless limit with non-zero cosmological constant.

On the other hand,
when we take a curved fiducial geometry, e.g.~the FLRW or an inhomogeneous background, it may be natural to promote it to a dynamical one.
In fact the dRGT massive gravity 
theory has been generalized to 
such a theory with two dynamical metrics, thereby called a bigravity theory, which is still free from the BD ghost \cite{HassanRosen}.
It contains a massless spin-2 field and a massive spin-2 field, with total seven degrees of freedom in the gravity sector \cite{deser}.

Phenomenologically, the bigravity theory has
many interesting features.
One of the biggest mysterious problems in modern cosmology
is the origin of the current accelerating expansion of the Universe \cite{IaSN}.
Based on the bigravity theory,  self-acceleration in 
all types
of the FLRW universe is allowed
\cite{Cosmology1,Cosmology2,growth, Cosmology3,Cosmology4,Cosmology5, with_twin_matter}.
Therefore, the present acceleration of the Universe 
can be explained by adding a mass to the graviton
(see also \cite{MassiveCosmo1, Gumrukcuoglu:2011ew, Gumrukcuoglu:2011zh, MassiveCosmo3, MassiveCosmo4, MassiveCosmo5,MassiveCosmo6, MassiveCosmo7}
about the cosmology in the dRGT theory).
Furthermore, since the bigravity theory contains two metrics,
a matter field in the fiducial metric sector 
is naturally introduced,
and it  may act as a dark matter component in the physical sector
\cite{with_twin_matter,bigravity_dark_matter}. 

However, although the background dynamics in the bigravity cosmology is viable,
a problem still exists at the perturbation level \cite{cp_instability1,cp_instability2,cp_instability3,cp_instability4}.
In the late stage of the Universe, 
the perturbations around the cosmological backgrounds are stable.
However, the perturbations suffer from either the ghost or gradient instability
in the sub-horizon scale,
when the Hubble expansion parameter is larger than the effective graviton mass.

Since the bigravity theory contains a massless spin-2 field,
the bigravity theory always contains GR solutions exactly as a special case.
If the two background metrics are
proportional, which we call a homothetic solution
\cite{Anisotropic2}, 
the basic equations are reduced to two sets of the Einstein equations in GR with a cosmological constant originated from the mass term.
That is, in homothetic spacetimes, 
the bigravity reproduces the background evolution of the universe identical to the one in GR.
The linear perturbations around a homothetic solution
are easily decomposed into two eigenstates: the massless and
massive graviton modes.
Here, the massive graviton mode is given by the FP theory
on a GR solution.
The perturbations in this mode suffer from the gradient instability
in the decelerating Friedmann universe,
while they have the Higuchi-type ghost
in the accelerating one
\cite{FP_on_FLRW}.
Therefore, the cosmological instabilities in the bigravity theory
are  similar to those in the FP theory on a cosmological background
and are related to the fact that 
the linear massive spin-2 field has no healthy massless limit
on the curved background.

However, such an instability is quite obscure in the physical interpretation,
since the natural expectation would be that
the massive theory should be restored to its corresponding massless theory
when the energy scale of the background spacetime is higher than the mass. 
Hence, the instability should be resolved without either a modification of the theory
or an extra ingredient,
if the bigravity theory is a reliable theory 
in such an energy scale.
The instability may simply hint the possibility that the linear perturbations are no longer valid.
Therefore, before we conclude the bigravity theory
breaks down in the early stage of the Universe,
it is instructive to
take into account the nonlinear interactions.

For this reason,
we will consider cosmological perturbations
with nonlinear effects in the bigravity theory.
As we will see in Sec.~\ref{sec_curved_background},
the instability arises from a scalar graviton mode.
Therefore, we will focus only on this mode and also restrict our analysis to a spherically symmetric configuration, for simplicity,
in which a gravitational degree of freedom is only given by the scalar graviton mode.

In order to find a stable model of the early Universe in the massive gravity 
theory (or the bigravity theory), there are several different approaches.
One possibility to obtain the viable cosmology
with a massive graviton is 
to extend the theory
\cite{quasidilaton, Gen_quasidilaton, rotation_massive, Generalized_massive}.
This approach may be justified by the reasoning that the bigravity or the dRGT theory may necessarily be
modified at a high energy scale
to realize a massive graviton resulting from spontaneous symmetry breaking,
and to protect the validity of the theory in the UV regime.
Another possibility is
to introduce
a doubly coupled matter field.
Although a matter field interacting with both metrics
suffers from a reappearing BD ghost in generic situations,
some ghost-free matter coupling was discussed
at a low-energy scale or in a homogeneous Friedmann background universe
\cite{Cosmology5, matter_coupling1,matter_coupling2,matter_coupling3,double_cosmo1, double_cosmo2, double_cosmo3,vielbein1,vielbein2,
Heisenberg1,Heisenberg2,Heisenberg3}.

In the present paper, however, we will show that
both the Higuchi-type ghost and the gradient instability
can be resolved by the nonlinear effects of the scalar graviton mode
in the ghost-free nonlinear bigravity theory
without either a modification of the theory or an additional matter field.
When the Hubble parameter is larger than the graviton mass,
we find the new cosmological solutions in the sub-horizon scale
for which the two spacetimes are still approximately homogeneous and isotropic,
and these two foliations
are related by a nonlinear coordinate transformation.
That is, the St\"{u}ckelberg fields  become non-linear,
while the spacetime perturbations are quite small.

The paper is organized as follows.
Introducing the ghost-free nonlinear bigravity theory,
we show the bigravity theory contains GR solutions as a special case
in Sec.~\ref{HR}.
In Sec.~\ref{linearized},
we derive the quadratic action around the GR solution,
which can be decomposed into  massless 
and massive graviton modes.
We discuss that the FP theory has an instability
on the cosmological background.
In Sec.~\ref{perturbation_around_FLRW}, retaining nonlinearities of the St\"uckelberg fields,
we study a spherically symmetric perturbation around 
the flat homothetic FLRW background
without matter perturbations,
and we discuss the stability in the early stage of the Universe.
Both Higuchi ghost and gradient instability
can be resolved by the nonlinear effects of the St\"uckelberg fields
in the early stage of the Universe.
In Sec.~\ref{cosmology_with_nonlinear},
we introduce the matter perturbation
and study  stability of the matter perturbations
both in the early stage and the late stage of the Universe.
We discuss  a possible transition from GR phase to the bigravity phase 
in the period when the graviton mass is comparable to the Hubble parameter
in Sec.~\ref{Sec_transition}.
We summarize our results and give some remarks in Sec.~\ref{summary}.
In Appendix \ref{flat_Vainshtein},
we present the Vainshtein screening in 
static spherically symmetric configurations
with a cosmological constant as well as twin matters.

\section{Nonlinear Bigravity Theory}\label{HR}
\subsection{Hassan-Rosen bigravity model}
In the present paper, we focus on the ghost-free 
bigravity theory proposed by Hassan and Rosen \cite{HassanRosen}, whose  action is given by
\begin{eqnarray}
\!\!\!\!\!\!\!\!\!\!  S &=&\frac{1}{2 \kappa _g^2} \int d^4x \sqrt{-g}R(g)+ \frac{1}{2 \kappa _f^2}
 \int d^4x \sqrt{-f} \mathcal{R}(f) \nonumber \\
&+&
S^{[\text{m}]}(g,f, \psi_g, \psi_f)
-\frac{m^2}{ \kappa ^2} \int d^4x \sqrt{-g} \mathscr{U}(g,f) 
\,,
\label{action}
\end{eqnarray}
where $g_{\mu\nu}$ and $f_{\mu\nu}$ are two dynamical metrics, and
$R(g)$ and $\mathcal{R}(f)$ are their Ricci scalars, respectively.
The parameters  $\kappa_g^2=8\pi G$ and $\kappa_f^2=8\pi \mathcal{G}$ are 
the corresponding gravitational constants, 
while $\kappa$ is defined by $\kappa^2=\kappa_g^2+\kappa_f^2$. 
We assume that the matter action $S^{[\text{m}]}$ 
is divided into two parts:
\bea
S^{[\text{m}]}(g,f, \psi_g, \psi_f)
=S_g^{[\text{m}]}(g,\psi_g)+S_f^{[\text{m}]}(f,\psi_f)
\,,
\ena
i.e.,  matter fields  $\psi_g$ and $\psi_f$ are coupled only to the $g$-metric 
and to the $f$-metric, respectively.
This restriction guarantees  
the weak equivalence principle as well as the ghost-free condition.
The $g$-matter $\psi_g$  and the $f$-matter $\psi_f$ are coupled gravitationally only 
through the interaction between two metrics $g$ and $f$.
We call $\psi_g$ and $\psi_f$  twin matter fluids \cite{Bimond}.

 The ghost-free interaction term between the two metrics,
often called the dRGT potential, is given by
\begin{equation}
\mathscr{U}(g,f)=\sum^4_{k=0}b_k\mathscr{U}_k(\gamma)
\,,
\end{equation}
{\setlength\arraycolsep{2pt}\begin{eqnarray}
&&\mathscr{U}_0(\gamma)=-\frac{1}{4!}\epsilon_{\mu\nu\rho\sigma} 
\epsilon^{\mu\nu\rho\sigma}\,, \nonumber \\
&&\mathscr{U}_1(\gamma)=-\frac{1}{3!}\epsilon_{\mu\nu\rho\sigma} 
\epsilon^{\alpha\nu\rho\sigma}
{\gamma ^{\mu}}_{\alpha}\,, \nonumber \\
&&\mathscr{U}_2(\gamma)=-\frac{1}{4}\epsilon_{\mu\nu\rho\sigma} 
\epsilon^{\alpha\beta\rho\sigma}
{\gamma ^{\mu}}_{\alpha}{\gamma ^{\nu}}_{\beta}\,, \\
&&\mathscr{U}_3(\gamma)=-\frac{1}{3!}\epsilon_{\mu\nu\rho\sigma} 
\epsilon^{\alpha\beta\gamma\sigma}
{\gamma ^{\mu}}_{\alpha}{\gamma ^{\nu}}_{\beta}{\gamma ^{\rho}}_{\gamma}\,, 
\nonumber \\
&&\mathscr{U}_4(\gamma)=-\frac{1}{4!}\epsilon_{\mu\nu\rho\sigma} 
\epsilon^{\alpha\beta\gamma\delta}
{\gamma ^{\mu}}_{\alpha}{\gamma ^{\nu}}_{\beta}{\gamma ^{\rho}}_{\gamma}
{\gamma ^{\sigma}}_{\delta}\,,
\nonumber
\end{eqnarray}}
where $b_k$ are coupling constants, while ${\gamma^{\mu}}_{\nu}$ is 
defined by 
\begin{equation}
{\gamma^{\mu}}_{\rho}{\gamma^{\rho}}_{\nu}
=g^{\mu\rho}f_{\rho\nu}
\,. 
\label{gamma2_metric}
\end{equation}

Taking the variation of the action with respect to $g_{\mu\nu}$ and
$f_{\mu\nu}$, we find two sets of the Einstein equations:
{\setlength\arraycolsep{2pt}\begin{eqnarray}
{G ^{\mu}}_{\nu} &=&
\kappa _g^2 ( {T ^{ [\gamma ] \mu} }_{\nu} 
+ {T^{\text{[m]} \mu} }_{\nu} ) \label{g-equation}, \\
{ \mathcal{G} ^{\mu}}_{\nu} &=& \kappa _f^2
( {\mathcal{T} ^{ [\gamma ] \mu} }_{\nu} 
+ {\mathcal{T}^{\text{[m]} \mu} }_{\nu} ), \label{f-equation}
\end{eqnarray}}
where ${G ^{\mu}}_{\nu}$ and ${ \mathcal{G} ^{\mu}}_{\nu} $ are the Einstein 
tensors for $g_{\mu\nu}$ and $f_{\mu\nu}$, respectively. 
The matter energy-momentum tensors 
${T ^{ [\rm m ] \mu} }_{\nu}$ and ${\mathcal{T} ^{ [\rm m ] \mu} }_{\nu}$
are given by
the variation of matter actions,
and
the $\gamma$-``energy-momentum" tensors 
${T ^{ [\gamma ] \mu} }_{\nu}$ and ${\mathcal{T} ^{ [\gamma ] \mu} }_{\nu}$
 are obtained by
the variation of the dRGT potential action
with respect to $g_{\mu\nu}$ and $f_{\mu\nu}$, respectively, 
taking the form \cite{bigravity_dark_matter,with_twin_matter}
\begin{equation}
{T^{[\gamma] \mu}}_\nu = \frac{m^2}{\kappa^2} \left( {\tau^\mu}_\nu - {\mathscr U} {\delta^\mu}_\nu \right) \; , \quad
{{\cal T}^{[\gamma] \mu}}_\nu = - \frac{\sqrt{-g}}{\sqrt{-f}} \frac{m^2}{\kappa^2} {\tau^\mu}_\nu
\end{equation}
where ${\cal \tau^\mu}_\nu \equiv \sum_{n=1}^4 \left( -1 \right)^{n+1} {\left( \gamma^{n} \right)^\mu}_\nu \sum_{k=0}^{4-n} b_{n+k} {\mathscr U}_k$.

The energy-momenta of matter fields are assumed to be 
conserved individually as
\begin{equation}
\overset{(g)}{\nabla} _{\mu}{T^{ [\text{m}] \mu} }_{\nu}=0\,,\;
\overset{(f)}{\nabla} _{\mu} {\mathcal{T} ^{ [\text{m} ] \mu} }_{\nu} =0
\,, 
\label{c1}
\end{equation}  
where $\overset{(g)}{\nabla} _{\mu}$ and $\overset{(f)}{\nabla} _{\mu}$ are 
covariant derivatives with respect to $g_{\mu\nu}$ and $f_{\mu\nu}$. 
From the contracted Bianchi identities for 
\eqref{g-equation} and \eqref{f-equation}, 
the conservation of the $\gamma$-``energy-momenta"  is
also guaranteed as
\begin{equation}
\overset{(g)}{\nabla} _{\mu}{T ^{ [\gamma] \mu} }_{\nu}=0\,,\;
\overset{(f)}{\nabla} _{\mu} {\mathcal{T} ^{ [\gamma] \mu} }_{\nu} =0
\,.
\label{c2}
\end{equation}
The Einstein equations \eqref{g-equation} and \eqref{f-equation}, together with the continuity equations \eqref{c1} and \eqref{c2}, determine the dynamics of the bigravity system given in \eqref{action}. In particular, \eqref{c2} are absent in GR, and in the massive/bigravity theory they give non-trivial constraints on cosmological solutions.

\subsection{Homothetic solution}
\label{homothetic}
First we give one simple set of solutions,
in which we assume that two metrics are proportional;
\bea
f_{\mu\nu}=K^2\, g_{\mu\nu}
\,,
\ena
where $K$ is a scalar function.
This ansatz provides a subset of the complete set of solutions, and
from the energy-momentum conservation (\ref{c2}), 
we find that $K$ is a constant.
As a result, we find two sets of the Einstein
equations with cosmological constants
$\Lambda_g$ and $\Lambda_f$:
\bea
G_{\mu\nu}(g)+\Lambda_g\,g_{\mu\nu}&=&\kappa_g^2 {T^{\text{[m]}}}_{\mu\nu}\,,
\label{homothetic_g}
\\
\mathcal{G}_{\mu\nu}(f)+\Lambda_f\,f_{\mu\nu} &=& \kappa _f^2
 {\mathcal{T}^{\text{[m]} } }_{\mu\nu} 
\label{homothetic_f}
\,,
\ena
where
\begin{align}
\Lambda_g(K)&=m^2\frac{\kappa_g^2}{\kappa^2}\,
\left(b_0+3b_1 K+3b_2 K^2+b_3 K^3\right)\,,
\nn
\Lambda_f(K)&=m^2\frac{\kappa_f^2}{\kappa^2}
\,\left(b_4+3b_3 K^{-1}+3b_2 K^{-2}+b_1K^{-3} \right)
\,. \label{eff_cc}
\end{align}
Since two metrics are proportional, we have the constraints on 
the cosmological constants and matter fields as
\bea
\Lambda_g(K)&=&K^2\Lambda_f(K)\,,
\label{eq_K}
\\
\kappa_g^2\,{T^{\text{[m]} \mu}}_{\nu}
&=&K^2 \kappa_f^2{\mathcal{T}^{\text{[m]} \mu} }_{\nu}
\,.
\label{matter_relation}
\ena
The quartic equation (\ref{eq_K}) fixes the value of $K$. 
It gives  at most four real roots, each of which gives a different cosmological constant.
The basic equations  (\ref{homothetic_g}) 
(or (\ref{homothetic_f})) are just the Einstein equations 
in GR with a cosmological constant.
Hence any solutions in GR with a cosmological constant are always
the solutions in the present bigravity theory.
We shall call these solutions homothetic solutions because of
the proportionality of the two metrics.



\section{Linear stability analysis of a homothetic spacetime}
\label{linearized}
\subsection{The perturbations around a homothetic solution}
\label{perturbation_homothetic}
The bigravity theory contains both a massless and massive spin-2 fields.
This becomes clear when we look at the linear perturbations around a homothetic solution.

The unperturbed solution is assumed to be  homothetic, i.e.,
\bea
\fB_{\mu\nu}=K^2  \gB_{\mu\nu}
\,,
\ena
where bar indicates background quantities. This provides solutions to the unperturbed part of the two Einstein equations \eqref{homothetic_g} and \eqref{homothetic_f}.
A constant $K$ is determined by the quartic equation (\ref{eq_K}), 
and the matter energy-momenta satisfy the
unperturbed part of \eqref{matter_relation}.

We then consider the following perturbations:
\begin{align}
g_{\mu\nu}&=\gB{}_{\mu\nu}+h^{[g]}_{\mu\nu},\\
f_{\mu\nu}&=\fB{}_{\mu\nu}+ K^2h^{[f]}_{\mu\nu}
=K^2\left(\gB{}_{\mu\nu}+h^{[f]}_{\mu\nu}\right)
\end{align}
where $|h^{[g]}_{\mu\nu}|, |h^{[f]}_{\mu\nu}| \ll |\gB{}_{\mu\nu}|$.
The suffixes of $h^{[g]}_{\mu\nu}$ 
as well as  $h^{[f]}_{\mu\nu}$ are raised and lowered by the background metric
$\gB_{\mu\nu}$.

We obtain the quadratic action
for the perturbations of the metrics as,
disregarding the tadpole-like term,
\begin{align}
S_2&=\int d^4x \sqrt{-\bar{g}}\Biggl[ 
\frac{1}{\kappa_g^2}\mathcal{L}_{\rm EH}\big[h^{[g]} ; \Lambda_g \big]+
\frac{K^2}{\kappa_f^2}\mathcal{L}_{\rm EH} \big[h^{[f]}  ; \Lambda_g\big]
\nn
&\qquad\qquad\qquad\quad
+\frac{1}{\kappa_-^2}\mathcal{L}_{\rm FP} \big[h^{[-]}  ; m_{\rm eff}^2 \big] 
 \Biggl] \nn
 &=\frac{1}{\kappa_+^2}\int d^4x \sqrt{-\bar{g}} \mathcal{L}_{\rm EH}\big[h^{[+]}  ; 
\Lambda_g\big]
 \nn
 &+\frac{1}{\kappa_-^2}\int d^4x \sqrt{-\bar{g}} 
\left[\mathcal{L}_{\rm EH}\big[h^{[-]}  ; \Lambda_g \big]
+\mathcal{L}_{\rm FP}\big[h^{[-]} ; m_{\rm eff}^2 \big]\right]\,,
\label{quad-act}
\end{align}
where the massless and massive graviton modes are defined by
\begin{align}
h^{[-]}_{\mu\nu}&=h^{[g]}_{\mu\nu}-h^{[f]}_{\mu\nu}\,, \\
h^{[+]}_{\mu\nu}&=\frac{\kappa_f^2}{K^2\kappa_-^2}h^{[g]}_{\mu\nu}
+\frac{\kappa_g^2}{\kappa_-^2}h^{[f]}_{\mu\nu}\,,
\end{align}
and $\kappa_+^2=K^{-2}\kappa_g^2\kappa_f^2/\kappa_-^2\,, \kappa_-^2
=\kappa_g^2+K^{-2}\kappa_f^2$
are the effective gravitational constants for the massless 
 and  massive graviton modes,
respectively.
The effective graviton mass in a homothetic background spacetime 
is defined by
\begin{align}
m_{\rm eff}^2&:=\frac{m^2}{\kappa^2}
\left(\kappa_g^2+\frac{\kappa_f^2}{K^2}\right)
(b_1K+2b_2K^2+b_3 K^3)
\,.\label{eff_mass}
\end{align}

The quadratic Einstein-Hilbert Lagrangian and the FP mass term
for a metric perturbation $h_{\mu\nu}$ are defined by
\begin{align}
\mathcal{L}_{\rm EH}[h  ; \Lambda_g]&=
-\frac{1}{4}h^{\mu\nu} \mathcal{E}_{\mu\nu,\alpha\beta}h^{\alpha\beta}
-\frac{\Lambda_g}{4}\left( h^2-2h_{\mu\nu}h^{\mu\nu}\right)\, , \\
\mathcal{L}_{\rm FP}[h ; m_{\rm eff}^2]&=-\frac{m_{\rm eff}^2}{8}\left( h_{\mu\nu}h^{\mu\nu}-h^2\right) \,,
\end{align}
where
\begin{align}
\mathcal{E}_{\mu\nu,\alpha\beta}h^{\alpha\beta}
&=-\frac{1}{2}\BoxB h_{\mu\nu}-\frac{1}{2}\nablaB_{\mu}\nablaB_{\nu}h
+\nablaB_{\alpha}\nablaB_{(\nu}h^{\alpha}{}_{\mu)}
\nn
&\quad
+\frac{1}{2}\bar{g}_{\mu\nu}\left( \BoxB h-\nablaB_{\alpha}\nablaB_{\beta}h^{\alpha\beta}\right)
\nn
&\quad
-2\left( h^{\alpha}{}_{(\mu}\bar{R}_{\nu)\alpha}-\frac{1}{2}h \bar{R}_{\mu\nu}\right)
\nn
&\quad
-\frac{1}{4}\left(\bar{g}_{\mu\nu}h-2h_{\mu\nu}\right)\bar{R}\,.
\end{align}
In the present paper, we assume the graviton mass 
is independent of the cosmological constant, although 
both of them are fixed by giving the coupling constants $\{b_k\}$ $(k=0 \dots 4)$ and 
the gravitational constants $\kappa_g^2$ and $\kappa_f^2$.
We treat the graviton mass and the cosmological constant 
as free parameters.
It is now explicit from \eqref{quad-act} that the bigravity theory contains one massless and one massive graviton.

\subsection{Massive graviton mode in a curved homothetic background spacetime}
\label{sec_curved_background}
At the linear order, the massive graviton mode in the bigravity theory
is given by the FP theory on  curved background.
We note that our background spacetime is dynamical 
unlike the original FP theory.

The massive mode $h^{[-]}_{\mu\nu}$ does not have a gauge symmetry,
since the mass term breaks the diffeomorphism.
The gauge symmetry can be explicitly restored
by introducing the  St\"uckelberg fields ${\cal A}_{\mu}$ and $\pi$ as
\begin{align}
h^{[-]}_{\mu\nu} = \mathcal{H}_{\mu\nu}+2\nablaB_{(\mu}{\cal A}_{\nu)}
+2\nablaB_{\mu}\nablaB_{\nu}\pi\,.
\end{align}
The perturbation $h^{[-]}_{\mu\nu}$ is invariant under the following 
gauge transformations:
\begin{align}
\mathcal{H}_{\mu\nu}\rightarrow \mathcal{H}_{\mu\nu}+2\nablaB_{(\mu}\xi_{\nu)}\,,
\quad
{\cal A}_{\mu}\rightarrow {\cal A}_{\mu}-\xi_{\mu}\,,
\end{align}
and
\begin{align}
{\cal A}_{\mu}\rightarrow {\cal A}_{\mu}+\nablaB_{\mu}\chi\,,
\quad
\pi\rightarrow \pi-\chi\,.
\end{align}
We can interpret $\mathcal{H}_{\mu\nu}$, ${\cal A}_{\mu}$, and $\pi$ 
 as tensor,  vector, and scalar graviton modes, respectively.

The Einstein-Hilbert action preserves the diffeomorphism invariance, and therefore neither ${\cal A}_\mu$ nor $\pi$ appears in $S_{\rm EH} \big[ h^{[-]}_{\mu\nu} \big]$. On the other hand,
the FP mass term is rewritten by using the St\"uckelberg fields as
\begin{align}
\mathcal{L}_{\rm FP}=
&-\frac{m_{\rm eff}^2}{8}(\mathcal{H}_{\mu\nu}\mathcal{H}^{\mu\nu}-\mathcal{H}^2)
-\frac{m_{\rm eff}^2}{8}{\cal F}_{\mu\nu}{\cal F}^{\mu\nu}
\nn
&+\frac{m_{\rm eff}^2}{2}\RB_{\mu\nu}{\cal A}^{\mu}{\cal A}^{\nu} 
-\frac{m_{\rm eff}^2}{2}(\mathcal{H}^{\mu\nu}\nablaB_{\mu}{\cal A}_{\nu}
-\mathcal{H} \nablaB_{\mu}{\cal A}^{\mu}) 
\nn
&+\frac{m_{\rm eff}^2}{2}\RB^{\mu\nu}\nablaB_{\mu}\pi\nablaB_{\nu}\pi
+m_{\rm eff}^2\RB_{\mu\nu}{\cal A}^{\mu}\nablaB^{\nu}\pi \nn
&-\frac{m_{\rm eff}^2}{2}(\mathcal{H}^{\mu\nu}\nablaB_{\mu}\nablaB_{\nu}\pi
-\mathcal{H} \BoxB \pi)
\label{FP_with_St}
\end{align}
where ${\cal F}_{\mu\nu}=2\nablaB_{[\mu}{\cal A}_{\nu]}$.
Note that the interaction terms 
between the tensor and the scalar graviton modes
produce the van Dam-Veltman-Zakharov (vDVZ)
 discontinuity around the flat background \cite{vDVZ}.
On the other hand, the vDVZ discontinuity does not occur
around a curved background in the massless limit \cite{no_vDVZ}.

In order to see the absence of the vDVZ discontinuity, 
we take a canonical normalization for the tensor and vector graviton modes as
\begin{align*}
\mathcal{H}_{\mu\nu}\rightarrow \kappa_-\mathcal{H}_{\mu\nu}, \quad
{\cal A}_{\mu}\rightarrow \frac{\kappa_-}{m_{\rm eff}}{\cal A}_{\mu}
\end{align*}
and for the scalar graviton mode as 
\begin{align*}
\pi\rightarrow \frac{\kappa_-}{m_{\rm eff}\sqrt{\RB_0}}\pi\,,
\end{align*}
where a positive constant $\RB_0$ denotes a typical scale of the background Ricci tensor,
i.e. $\RB_0 \sim O(\RB_{\mu\nu})$.
Then the scalar graviton part of the action becomes
\begin{align}
S_2 \supset \int d^4x \sqrt{-\bar{g}} \Biggl[
& \frac{\RB^{\mu\nu}}{2\RB_0}\nablaB_{\mu}\pi\nablaB_{\nu}\pi
+\frac{\RB_{\mu\nu}}{\sqrt{\RB_0} }{\cal A}^{\mu}\nablaB^{\nu}\pi \nn
& -\frac{m_{\rm eff}}{2\sqrt{\RB_0} }(\mathcal{H}^{\mu\nu}\nablaB_{\mu}\nablaB_{\nu}\pi
-\mathcal{H} \BoxB \pi)
\Biggl]\,.
\end{align}
Therefore, when the effective graviton mass is negligible compared with the
background Ricci curvature (i.e. $m_{\rm eff}^2 \ll \RB_0$),
the interaction between tensor and scalar modes vanishes,
and then there is no vDVZ discontinuity. 

However the kinetic term of the scalar graviton is modified from 
the standard one in such a massless limit.
Hence the ghost instability or the gradient instability may appear,
depending on the background Ricci curvature.
For instance, we consider the background spacetime is given by
the flat FLRW universe:
\begin{align}
d\bar{s}_g^2=a^2(-d\eta^2 +d{\bf x}^2)\,.
\end{align}
The kinetic term is expressed by
\begin{align}
\RB^{\mu\nu}\nablaB_{\mu}\pi\nablaB_{\nu}\pi
=\frac{3H^2}{2a^2}(1+3w)\left[(\partial_{\eta}\pi)^2-\frac{w-1}{1+3w}(\partial_i \pi)^2 \right]
\,,
\label{kinetic_curved}
\end{align}
where  $w$ is the effective equation-of-state parameter of the background universe
defined later by \eqref{def_w}, and
$H=\dot{a}/a^2$ is the Hubble parameter where a dot is
the derivative with respect to the conformal time $\eta$.  
Therefore the Higuchi ghost type instability appears for $w<-1/3$, 
while the gradient instability is found for $-1/3<w<1$
\cite{FP_on_FLRW}.
This fact indicates that an instability is unavoidable 
if the background spacetime consists of an ordinary matter components
$(w<1)$.

It was shown that the gradient instability appears in a more general
non-homothetic but homogeneous and isotropic background in bigravity
\cite{cp_instability1,cp_instability2,cp_instability3}.

The instabilities indicate that the linear approximation is no longer valid 
in such a background universe.
Although the interaction between tensor and scalar modes
is suppressed by the small coefficient $m_{\rm eff}/\sqrt{\RB_0}$,
the interaction between the tensor mode and the scalar mode 
cannot be ignored
because the unstable scalar mode grows exponentially in time.
As a result, unless this instability is resolved,
the fifth force will be recovered, and then the vDVZ discontinuity will reappear.


Especially when we discuss the early stage of the Universe,
in which we hope to recover a standard big bang universe, 
this instability becomes a serious problem.
If the instability and vDVZ discontinuity are not resolved,
 the bigravity theory 
cannot realize the realistic homogeneous and isotropic universe
without an elaborate fine-tuned initial condition.

The above argument is only based on the linear theory.
The existence of the instability indicates that
nonlinear interactions cannot be ignored
\footnote{ Some nonlinear interactions by introducing the St\"uckelberg fields
on the curved background
are discussed in \cite{nonlinear_stu}.}.
This instability is supposed to arise from the scalar graviton mode.
Hence, the nonlinear interactions of the scalar graviton mode
must be taken into account.
If the above linear instability is stabilized and then the small 
coefficient proportional to the graviton mass term is kept to be small 
enough, the standard big bang universe can be recovered as a stable solution 
in the early stage of the Universe.
It is similar to the Vainshtein screening mechanism in which
the nonlinear effects of the scalar graviton are essential.
We shall call our case the cosmological Vainshtein mechanism.

For this reason, we consider the perturbation
around the flat FLRW background
retaining nonlinearities of the St\"uckelberg fields.
We then discuss whether the St\"uckelberg fields
can be stabilized by nonlinear interactions of the scalar graviton,
and whether the fifth force can be screened in the early stage of the Universe.


\section{Scalar graviton with nonlinear effects}
\label{perturbation_around_FLRW}
\subsection{Strategy}

To ease the difficulty in analyzing
non-linear effects for a generic spacetime, we restrict our analysis to a spherically symmetric configuration 
of cosmological solutions.
Even in a spherically symmetric system, however,
it is still difficult to discuss
full nonlinear effects without resorting to numerical simulations.
Hence, we consider some simplified case.
In this subsection, we summarize the strategy for our analysis.

We consider non-linear perturbations on homothetic flat FLRW backgrounds:
\begin{align}
d\bar{s}_g^2&=a^2(\eta)(-d\eta^2+dr^2+r^2d\Omega^2)\,,\\
d\bar{s}_f^2&=K^2a^2(\eta)(-d\eta^2+dr^2+r^2d\Omega^2)\,.
\end{align}
where $d\Omega^2=d\theta^2+\sin^2 \theta d\varphi^2$.
This homothetic solution satisfies
\begin{align}
&3H^2=\kappa_g^2\bar{\rho}_g+\Lambda_g ,
\label{Friedmann_background}
\\
&
\frac{\dot{\bar{\rho}}_g}{a}+3H(\bar{\rho}_g+\bar{P}_g)=0\,,
\label{unperturbed}
\end{align}
with 
\begin{align}
&\Lambda_g=K^2\Lambda_f
,\,\,
\label{eq_K2}
\\
&
\kappa_g^2\bar{\rho}_g=K^2\kappa_f^2\bar{\rho}_f
\; , \quad
\kappa_g^2\bar{P}_g=K^2\kappa_f^2\bar{P}_f
\label{eq_matter}
\,,
\end{align}
where $H=\dot{a}/a^2$ and a dot is the derivative with respect to 
the conformal time $\eta$.
We define the effective equation-of-state parameter $w$ by
\begin{align}
w:=\frac{\kappa_g^2\bar{P}_g-\Lambda_g}{\kappa_g^2\bar{\rho}_g+\Lambda_g}
=-1-\frac{2\dot{H}}{3aH^2}
\,.
\label{def_w}
\end{align}

For general non-linear perturbations, it is difficult, if not impossible, to 
do an analysis even for spherically symmetric system without numerical simulations.
Hence we discuss the following approximated model.
First we impose spherical symmetry at full order and
assume that the $g$- and $f$-spacetimes are approximated by the FLRW metric
such that 
\begin{align}
ds_g^2&=a^2(\eta_g)\left[
-e^{2\Phi_g}d\eta_g^2+e^{2\Psi_g}dr_g^2+r_g^2d\Omega^2\right],\\
ds_f^2&=K^2a^2(\eta_f)\left[
-e^{2\Phi_f}d\eta_f^2+e^{2\Psi_f}dr_f^2+r_f^2d\Omega^2\right]\,,
\label{}
\end{align}
where we introduce two coordinate systems $(\eta_g, r_g)$ and 
$(\eta_f, r_f)$ to describe the approximated FLRW spacetimes,  
which are given by different coordinate transformations 
from the original one coordinate system $(\eta, r)$ as
\begin{align*}
\eta_g&=\eta_g(\eta,r),\quad
r_g=r_g(\eta,r), \\
\eta_f&=\eta_f(\eta,r),\quad
r_f=r_f(\eta,r)\,.
\end{align*}
The approximated FLRW spacetimes mean that we assume 
 $|\Phi_g|,|\Psi_g| \ll 1$ and $|\Phi_f|,|\Psi_f| \ll 1$ because 
the mass interaction term, which is proportional to $m_{\rm eff}^2/R_0$ 
and gives the deviation from GR, is assumed be small.
However they do not mean $(\eta_g, r_g) \approx (\eta, r)$
 and $(\eta_f, r_f)\approx (\eta, r)$, 
in which case the deviation from homothetic spacetimes is small 
and then can be described by the linear perturbations.

Although the bigravity theory allows one coordinate transformation,
two independent coordinate 
transformations can be possible apparently by introducing the St\"uckelberg field
such that 
\begin{align}
\eta_f=\eta_g+{\cal A}^{\eta}\,,~~~r_f=r_g+{\cal A}^r\,,
\label{gauge-freedom}
\end{align}
where $({\cal A}^{\eta}, {\cal A}^r)$ is the St\"uckelberg field 
in the spherically symmetric case.
Using a gauge freedom, we can fix one coordinate system.
%

We also assume that for the unperturbed FLRW spacetimes,
$K$ and $a$ are given by Eqs. (\ref{unperturbed}), (\ref{eq_K2})
 and (\ref{eq_matter})
although they are not homothetic.
This is allowed as we obtain the consistent 
perturbation equations with this ansatz.

For the following discussions, 
we consider only  a sub-horizon scale $(aL\ll H^{-1})$ and 
a length smaller than the Compton wave length of 
the massive graviton $(aL \ll m_{\rm eff} ^{-1})$. 
This is because our interest is the sub-horizon physics during the epoch $H > m_{\rm eff}$ and
such a solution provides us a stable cosmological 
Vainshtein mechanism.
We define a dimensionless parameter as
\begin{align}
\epsilon:={aL\over H^{-1}}=aLH\,, 
\end{align}
which satisfies $\epsilon \ll 1$ for a sub-horizon scale.

For a spherically symmetric spacetime, 
we can divide the behaviour of all variables into two:
one is a slowly changing  longitudinal mode mainly due to the background 
expansion and matter distributions, and 
the other is a fast changing wave-like oscillation mode of a scalar graviton. 
If the wave amplitude of the oscillation mode is small, when we take
 an average over the typical scale of the system $aL$, 
which is smaller than the horizon scale $H^{-1}$,
we find only longitudinal-mode variables, which we call adiabatic modes.
We then decompose all variables $X$ into
adiabatic and oscillation modes as
\begin{align}
X=X^{\rm ad}+\chi^{\rm osc}\,,
\end{align} 
with 
\begin{align}
X^{\rm ad}\approx \left< X \right>\,,
\end{align}
where $\left< \;\; \right>$ denotes an average over the typical scale of the system.

The dynamical time scale of the adiabatic mode 
is assumed to be the Hubble time scale, and then 
its evolution is caused by the expansion of the Universe
and of the density perturbations.
On the other hand, the oscillation mode
comes from the degree of freedom of the scalar graviton.
The time scale of the oscillation mode may  be 
the same order of the inhomogeneity scale.
Then each change rate is evaluated as
\begin{align}
|\partial_{\eta} X^{\rm ad}| &\sim |aH X^{\rm ad}|
\,, 
\label{adiabatic_time_scale} \\
|\partial_{\eta} \chi^{\rm osc}|
&\sim
|\partial_{r}\chi^{\rm osc}|\,.
\end{align}
Since we consider a sub-horizon scale ($aL<H^{-1}$), 
the dynamical time scale of an oscillation mode
is much shorter than the Hubble expansion time, i.e., 
\begin{align}
|\partial_{r}\chi^{\rm osc}|\gg |aH \chi^{\rm osc}|\,. 
\end{align}

Since the adiabatic mode may be obtained 
by taking the spatial average over the typical scale of the system, 
if the oscillation mode is small,
we can assume that the dynamics of the adiabatic mode is decoupled
from the dynamics of the oscillation mode.
This assumption is valid if the small oscillation mode has no instability.
Hence we first consider the evolution of the adiabatic modes without 
the oscillation modes.
Then we study the dynamics of the oscillation modes around this adiabatic solution.


\subsection{Adiabatic mode solution}
In this subsection we discuss the time evolution of the adiabatic modes
for the case without matter perturbations  in order to see 
the behaviour of non-linear  St\"uckelberg field.
The full analysis including matter perturbations 
will be discussed in Sec. \ref{cosmology_with_nonlinear},
and the explicit expressions will be summarized in Sec \ref{adiabatic_with_matter}. 

For the adiabatic modes, we fix the gauge freedom \eqref{gauge-freedom} by setting 
\begin{align}
\eta_g=\eta \; ,\quad
r_g=r \; ,
\end{align}
and introduce the dimensionless variables $\nu$ and $\mu$ to parametrize $\eta_f$ and $r_f$ as
\begin{align}
\eta_f=(1+\nu)\eta,\quad
r_f=(1+\mu)r\,.
\end{align}
We assume that
the time coordinate $\eta_f$ and the radial coordinate $r_f$
point the same directions of $\eta_g$ and $r_g$, respectively,
i.e.,
\begin{align}
\nu>-1\,, \quad \mu>-1\,.
\end{align}
We have assumed the weak inhomogeneous gravitational fields around 
the FLRW spacetimes,
i.e.,
\begin{align}
|\Phi_g|,|\Psi_g|,|\Phi_f|,|\Psi_f| \ll 1\,,
\end{align}
and
\begin{align}
|r\Phi_g'|,|r\Psi_g'|,|r\Phi_f'|,|r\Psi_f'| \ll 1\,,
\end{align}
which means that the perturbations from homogeneous and isotropic spacetimes
are small. 
Note that this does not imply that the perturbations 
from the homothetic FLRW spacetime are small because of the existence of 
the non-linear St\"uckelberg field, i.e.,
either $\nu$ or $\mu$ are not necessarily small.
The nonlinearities in the variables $\nu$ and $\mu$ must be retained.
However we assume those variables 
are not so large such that the perturbations of gravitational fields are still small, i.e., 
\begin{align}
|\mu \Phi_g| \ll 1 \,, \quad
|r\mu' \Phi_g| \ll 1 \,,
\cdots\,.
\end{align}
The spatial derivative of the adiabatic mode may be evaluated by
\begin{align}
|\partial_r X^{\rm ad}| \sim L^{-1} |X^{\rm ad}|\,,
\quad L\lesssim r\,,
\label{adiabatic_spatial_scale}
\end{align}
which leads with  \eqref{adiabatic_time_scale}  to
\begin{align}
|\partial_{\eta} X^{\rm ad}| \sim \epsilon |\partial_r X^{\rm ad}|
\ll |\partial_r X^{\rm ad}| \,.
\label{adiabatic_assumption}
\end{align}

Since the dynamical time scale of the adiabatic mode variables
is given by $H^{-1}$, 
our spherically symmetric solution around the FLRW spacetimes must be 
restored to 
the static solution in the limit of $H\rightarrow 0$,
which is shown in Appendix \ref{flat_Vainshtein}
(see also \cite{Vainshtein}).

In a spherically symmetric static solution, 
 the metric $f_{\mu\nu}$  has no 
non-diagonal component in the coordinates $(\eta,r)$.
Hence 
the non-diagonal component of $f_{\mu\nu}$
is at most of the order of $\epsilon$ 
in the present adiabatic solution around the FLRW spacetimes.
The non-diagonal component is given by
\begin{align}
f_{\eta r}&=
- K^2
a^2[e^{2\Phi_f} (\eta+\eta\nu )^{\cdot}\eta\nu'-e^{2\Psi_f}(r+r \mu)'r \dot{\mu}]
\,.
\end{align}
where a prime denotes the derivative with respect to $r$.
Because $\dot{\nu}\sim aH\nu,\, \dot{\mu}\sim aH\mu$
and $\vert \Phi_f \vert , \vert \Psi_f \vert \ll 1$,
we find the leading contribution as,
assuming $K \sim {\cal O}(1)$,
\begin{equation}
f_{\eta r}  
\sim a^2 \, {\cal O}(\eta \nu' , \epsilon \mu , \epsilon \mu^2)
\end{equation}
which must be ${\cal O}(\epsilon)$.
Since $\eta \nu'\sim \nu/\epsilon\sim {\cal O}(\epsilon)$, 
  the  St\"uckelberg variables $\nu$ and $\mu$ are evaluated as
\begin{align}
|\nu| \lesssim {\cal O}(\epsilon^2)  
,\quad
|\mu| \lesssim {\cal O}(1)\,.
\label{nu_and_mu}
\end{align}

We expand all basic equations up to the second order of $\epsilon$.
The inhomogeneous gravitational fields $\Phi_g,\Psi_g,\Phi_f,\Psi_f$ are determined by 
the Einstein equations, whose explicit solutions 
are  given by Eqs.~\eqref{psi_g}-\eqref{phi_f} 
in Sec.~\ref{adiabatic_with_matter}.
The  St\"uckelberg variable $\nu$ is solved by 
the $\eta$ component of the interaction conservation law 
$\nabla_{\alpha}T^{[\gamma]\alpha}{}_{\beta}=0$
as
\begin{align}
\frac{\partial}{\partial r}\Biggl[ \frac{r^2}{2+(r\mu)'}&\Big(\eta \nu'+arH\mu(2+(r\mu)')+r\dot{\mu} \Big)
\nn
&\times (1+2\beta_2 \mu +\beta_3 \mu^2)\Biggl]=0\,,
\label{solve_nu}
\end{align}
where the parameters $\beta_2$ and $\beta_3$ are defined by
\begin{align}
\beta_2&:=\frac{b_2K^2+b_3K^3}{b_1K+2b_2K^2+b_3K^3}\,, 
\\
\beta_3&:=\frac{b_3K^3}{b_1K+2b_2K^2+b_3K^3}\,.
\end{align}
Eq.~\eqref{solve_nu} is integrable.
An integral constant must be zero because of 
the regularity condition at $r=0$.
Hence we obtain two cases:
The first parenthesis in the square brackets of \eqref{solve_nu} vanishes or the second one does so.
If the second parenthesis vanishes, 
$\mu$ is a constant. 
However, barring special tuning of model parameters, such a solution cannot reproduce
the static result in the limit of  $H\rightarrow 0$.
Hence we conclude that the first parenthesis vanishes, i.e.,
\begin{align}
\eta \nu'=-Har\mu(2+(r\mu)')-r\dot{\mu}\,,
\label{nu_eq}
\end{align}
which determines $\nu$ by giving $\mu$.
This expression shows that 
the condition \eqref{nu_and_mu} is consistent.

Substituting 
\eqref{nu_eq}, together with later obtained Eqs.~\eqref{psi_g}-\eqref{phi_f},
into the $r$ component of $\nabla_{\alpha}T^{[\gamma]\alpha}{}_{\beta}=0$,
we obtain an algebraic equation for another St\"uckelberg variable $\mu$ as
\begin{align}
&\mathcal{C}_{m^2}(\mu)+\mathcal{C}_{H^2}(\mu)=0 \label{mu_eq_without_matter}\,,
\end{align}
where both $\mathcal{C}_{m^2}$ and $\mathcal{C}_{H^2}$
are quintic functions of $\mu$
 (The explicit forms are given in Sec.~\ref{adiabatic_with_matter}).
These terms have typical magnitudes given by 
\begin{align*}
\mathcal{C}_{m^2}&\sim m_{\rm eff}^2 \times \mathcal{O}(\mu)
\,,\quad
\mathcal{C}_{H^2}\sim H^2 \times \mathcal{O}(\mu)
\,.
\end{align*}
The equation \eqref{mu_eq_without_matter} reproduces the static result \eqref{mu_eq2}
in the limit of $H\rightarrow 0$.

Since $\mu$ is determined by the algebraic equation \eqref{mu_eq_without_matter},
$\mu$ has no dynamical degree of freedom.
It is not surprising because
we have  ignored the oscillation mode 
which corresponds to the dynamical degree of freedom of the scalar graviton.
As a result, 
the St\"uckelberg fields do not have any dynamical freedom 
 in the adiabatic mode solutions.

From now on, we focus on the period of the Universe with 
 $H \gg m_{\rm eff}$, which corresponds to the early stage of the Universe.
The algebraic equation \eqref{mu_eq_without_matter} reduces
\begin{align}
\mathcal{C}_{H^2}\approx 0\,.
\label{algebraic_H>>m}
\end{align}
This equation has at most four roots, which are 
 given by $\mu=-1$ and
\begin{align}
\mu_0=0,~{\rm and}~~\mu_{\pm} \,,
\end{align}
where
\begin{widetext}
\begin{align}
\mu_{\pm} 
=
\frac{1+(1-3w)\beta_2 \pm \sqrt{1-4\beta_2+(1-3w)^2\beta_2^2+3(1-w)(1+3w)\beta_3}}{-2\beta_2+(1+3w)\beta_3} \; .
\label{mu_pm}
\end{align}
\end{widetext}
Since the root  $\mu=-1$ gives $r_f=0$ for any $r$,
we do not adopt this solution,
and consider only the other three roots $\mu_0 = 0 , \mu_\pm$.
Since those roots are constants, which depend on the coupling 
constants and equation-of-state parameter, 
we can classify the solutions of Eq. (\ref{mu_eq_without_matter}) by those roots, which 
we call the $\mu_0$-branches.

When we do not include matter perturbations,
neglecting the contributions from the interaction terms
(which are much smaller than $\epsilon^2$ for $H\gg m_{\rm eff}$),
the metric perturbations are solved as
\begin{align}
\Phi_g,\Psi_g,\Phi_f,\Psi_f \approx 0\,.
\end{align}
Then two metrics are given by
\begin{align}
ds_g^2&=a^2(\eta)\left[
-d\eta^2+dr^2+r^2d\Omega^2\right]\,,\nn
ds_f^2&=K^2a^2(\eta_f)\left[
-d\eta_f^2+dr_f^2+r_f^2d\Omega^2\right]\,,
\label{cos_sol}
\end{align}
where the coordinates $(\eta_f,r_f)$, which correspond to 
the nonlinear modulations due to the adiabatic
St\"uckelberg fields,
deviating from the physical coordinates as
\begin{align}
\eta_f=\eta-\frac{1}{2}Har^2(2\mu_0+\mu_0^2)
,\quad
r_f=(1+\mu_0)r
\,,
\end{align}
where we have integrated Eq.~\eqref{nu_eq} for $\nu$
setting $\mu=\mu_0$.

The  solution $\mu_0=0$ corresponds to the homothetic FLRW spacetimes,
while we can also find the other approximately homogeneous and isotropic solutions
with $\mu_0 = \mu_{\pm}$ in the massless limit,
in which the coordinate transformation from $(\eta,r)$ 
to $(\eta_f,r_f)$ is nonlinear.
The solutions with $\mu_0 = \mu_{\pm}$ are valid  
up to the second order of $\epsilon$
with $\epsilon \ll 1$
when the interaction terms can be ignored.
Thus the solutions with the nonlinear  St\"uckelberg variable $\mu$ 
are not exactly homogeneous and isotropic spacetimes,
but approximate homogeneity and isotropy still hold in the sub-horizon scales.


\subsection{Stability conditions of scalar graviton}
\label{sec_stability}

Next we consider the oscillation modes of perturbations.
In the previous subsection, we discussed a spherically symmetric solution
based on the adiabatic potential approximation.
It does not contain 
the dynamical degree of freedom of the scalar graviton.
In this subsection, we analyze 
the stability of the solution against the fluctuations of the scalar graviton
with the condition 
 $H\gg m_{\rm eff}$.
The stability for the case with 
 $H\ll m_{\rm eff}$ is discussed in Appendix \ref{app_late_stability}.

We consider the following perturbations:
\begin{align}
ds_g^2&=a^2(\eta_g)\left[
-e^{2(\Phi_g+\phi_g)}d\eta_g^2+e^{2(\Psi_g+\psi_g)}dr_g^2+r_g^2d\Omega^2\right],\\
ds_f^2&=K^2a^2(\eta_f)\left[
-e^{2(\Phi_f+\phi_f)}d\eta_f^2+e^{2(\Psi_f+\psi_f)}dr_f^2+r_f^2d\Omega^2\right]\,.
\end{align}
We divide the perturbations into the adiabatic and oscillation modes.
When we take an average over the typical scale of the system, the oscillation-mode perturbations
do not contribute, and the equations for the adiabatic equations are obtained.
We solved them in the previous subsection.
When the oscillation-mode perturbations are defined by 
\begin{align}
\chi^{\rm osc}=X-X^{\rm ad}
\,,
\end{align}
the equations that govern their evolution are found by subtraction of the adiabatic modes from 
the full perturbation equations.

Using a gauge freedom of the oscillation-mode perturbations, 
as in \eqref{gauge-freedom},
we set two coordinates as
\begin{align}
&\eta_g=\eta+\delta \eta(\eta,r)\,,\quad
r_g=r+\delta r(\eta,r)\,,\\
&\eta_f=\eta-\frac{1}{2}Har^2(2\mu_0+\mu_0^2)
\,,\quad
r_f=(1+\mu_0)r\,,
\end{align}
where we have used the previous solutions for the adiabatic mode.

While $(\Phi_g, \Psi_g, \Phi_f, \Psi_f)$ are the adiabatic modes,
$(\phi_g,\psi_g,\phi_f,\psi_f,\delta \eta,\delta r)$ are 
the oscillation modes of perturbations.
We assume that all oscillation-mode variables 
have small amplitudes, 
i.e., $|\chi^{\rm osc}| \ll 1$, 
and the rate of their change in time is roughly
 $|\partial_{\eta} \chi^{\rm osc}| \sim |\partial_r \chi^{\rm osc}|$.

We  find that the perturbed metric variables are not dynamical 
and they vanish in the limit of $m_{\rm eff}/H\rightarrow 0$.
This is easy to see from the equation of motion as follows:
The Einstein curvature tensors contain
the terms proportional to $H^2$,
while the energy-momentum tensors of the interaction term 
are proportional to $m^2_{\rm eff}$.
For instance, the $(\eta_g,\eta_g)$-component of the Einstein equations
in the coordinates $(\eta, r)$
gives
\begin{align}
&\quad
6H^2\phi_g-\frac{2}{a^2r^2}\frac{\partial(r\psi_g)}{\partial r}
-\frac{2H}{a}\frac{\partial \psi_g}{\partial \eta}
\nn
&
=m_{\rm eff}^2\frac{\kappa_g^2}{\kappa_-^2} \biggl[
{(1+2\beta_2\mu_0+\beta_3\mu_0^2)\over r^{2}}
\frac{\partial (r^2\delta r)}{\partial r}
+\cdots \biggl]\,.
\end{align}
Taking into account other components of the Einstein equations, we see that
the energy-momentum tensors of the interaction term 
are negligible compared to the Einstein tensors when 
\begin{align}
|\phi_g|,|\psi_g| \gg \frac{\kappa_g^2}{\kappa_-^2} 
\frac{m_{\rm eff}^2}{H^2} |\partial_a\delta r|,\;
\frac{\kappa_g^2}{\kappa_-^2} 
\frac{m_{\rm eff}^2}{H^2}  |\partial_a \delta \eta| \,,
\label{negligible2}
\end{align}
where $\partial_a =(\partial_\eta,\partial_r)$.
When the conditions \eqref{negligible2} hold,
i.e., for the early stage of the Universe 
with $m_{\rm eff} \ll H$,
it is justified that the Einstein equations in bigravity 
is restored to the GR form.
In such a stage,
the Einstein equations for $g_{\mu\nu}$ give the solution:
\begin{align}
\phi_g\approx 0
,\quad
\psi_g\approx 0
\,.
\end{align}
This result is convincing because 
there is no dynamical degree of freedom in a spherically symmetric system
without matter perturbations in GR.
By the same argument as above, we also find
\begin{align}
\phi_f\approx 0
,\quad
\psi_f\approx 0
\,,
\end{align}
from the Einstein equations for $f_{\mu\nu}$.

In the limit of $m_{\rm eff}/H\rightarrow 0$, we find that two metrics
without matter perturbations are given by
\begin{align}
ds_g^2&=a^2(\eta_g)\left[
-d\eta_g^2+dr_g^2+r_g^2d\Omega^2\right],\\
ds_f^2&=K^2a^2(\eta_f)\left[
-d\eta_f^2+dr_f^2+r_f^2d\Omega^2\right]\,.
\end{align}
We then expand the action in terms of $(\delta \eta, \delta r)$
up to the second order of $\epsilon$.
The variations 
with respect to $\delta \eta$ and $\delta r$
give the constraint equations,
which are solved such that the St\"uckelberg variables 
$\delta \eta$ and $\delta r$ are given by 
\begin{align}
\delta \eta&=-\frac{\partial_{\eta} \pi}{a^2} 
+ \frac{arH \mu_0}{1+\mu_0}\frac{\partial_r \pi}{a^2}
+\mathcal{O}(\epsilon^2)\,,
\label{delta_eta}\\
\delta r&=\frac{\partial_{r} \pi}{a^2(1+\mu_0)}
+\frac{arH \mu_0}{1+\mu_0}\frac{\partial_{\eta} \pi}{a^2}
+\mathcal{O}(\epsilon^2)\,,
\label{delta_r}
\end{align}
in terms of  a St\"uckelberg scalar  $\pi$.
For the analysis of stability, it is sufficient
to determine $\delta \eta$ and $\delta r$ up to 
the first order of $\epsilon$.

Substituting \eqref{delta_eta} and \eqref{delta_r}
into the action,
we obtain the quadratic action of $\pi$ as
\begin{align}
S_2&=\frac{m_{\rm eff}^2}{\kappa_-^2}
\int d\Omega  
\int d\eta dr (arH)^2 
\mathcal{K}_S \left[\left(\partial_{\eta}\pi\right)^2
-c_S^2\left(\partial_{r}\pi\right)^2 \right] \; ,
\end{align}
where $\Omega$ is the solid angle.
The signs of these coefficients $\mathcal{K}_S$ and $c_S^2$ determine 
the stability of the St\"uckelberg scalar $\pi$, which corresponds to 
the scalar graviton. 
 
For  the $\mu_0=0$\,-\,branch, the coefficients are given by 
\begin{align}
\mathcal{K}_S|_{\mu_0=0}&=\frac{3}{4}(1+3w)\,,
\\
c_S^2|_{\mu_0=0}&=\frac{w-1}{1+3w}\,,
\end{align}
which is consistent with the result \eqref{kinetic_curved}.
On the other hand, for  the $\mu_0=\mu_{\pm}$\,-\,branches,
after simplifying the expressions by using \eqref{mu_pm},
we find
\begin{align}
\mathcal{K}_S|_{\mu_0=\mu_{\pm}}&=\frac{3}{4}(3w-1)(2+\mu_{\pm})
(1+2\beta_2\mu_{\pm}+\beta_3\mu_{\pm}^2)\,,
\nn
&~\\
c_S^2|_{\mu_0=\mu_{\pm}}&=\frac{2\left(3(1-w)+(1-(3w-1)\beta_2)\mu_{\pm}\right)}
{3(3w-1)(2+\mu_{\pm})(1+2\beta_2\mu_{\pm}+\beta_3\mu_{\pm}^2)}\,.\nn
&~
\end{align}
The no-ghost condition is given by 
$\mathcal{K}_S>0$ while  
the no-gradient instability condition is 
given by $c_S^2>0$.
Hence, the stability condition of the scalar graviton is to have
\begin{align}
\mathcal{K}_S>0
,\quad
c_S^2>0\,.
\label{no_gg}
\end{align}

\begin{figure*}[htb]
\centering
\includegraphics[width=16cm,angle=0,clip]{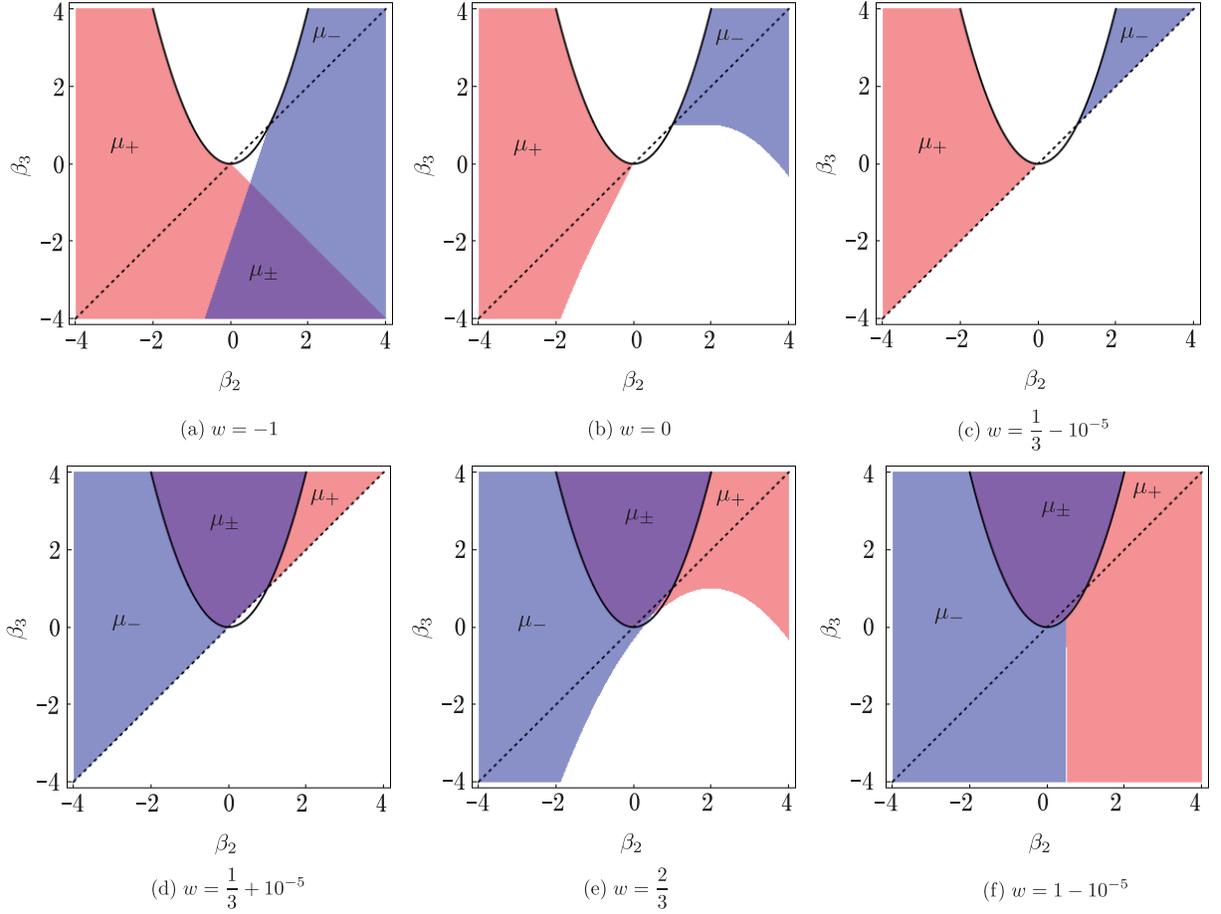}
\caption{The stable regions of the coupling constants
for flat FLRW backgrounds
for the cases of $w=-1$, $w=0$, $w=1/3\pm 10^{-5}$, $w=2/3$ and $w=1-10^{-5}$. 
The  $\mu_0=\mu_+$\,-\,branch
is stable in the regions denoted by $\mu_+$ (red regions),
the $\mu_0=\mu_-$\,-\,branch
 is stable in the regions denoted by $\mu_-$ (blue regions),
and
both $\mu_0=\mu_{\pm}$ are stable in the regions denoted by $\mu_{\pm}$ 
(purple regions).
The  black solid curves correspond to $\beta_3=\beta_2^2$
and the black dotted lines correspond to $\beta_2=\beta_3$.
}
\label{no_instability1}
\end{figure*}

In the $\mu_0=0$\,-\,branch,
either the ghost instability or the gradient instability
appears for $w<1$.
So the instability is 
inevitable when the Universe consists of the standard matter,
as shown in Sec.~\ref{sec_curved_background}.
Since the instability appears only in the 
relation between the two coordinate systems, $(\eta_g,r_g)$ and $(\eta_f,r_f)$,
two spacetimes still keep homogeneous approximately
as long as the condition \eqref{negligible2} holds.
However, since $\pi$ grows in time due to  the instability in this branch, 
the condition \eqref{negligible2}
eventually breaks down. 

On the other hand, 
the  $\mu_0=\mu_{\pm}$\,-\,branches 
can avoid the ghost instability as well as the gradient instability
depending on the background dynamics and the coupling constants.
In Fig.~\ref{no_instability1}, 
we show the parameter regions where the solution is stable
for the cases of $w=-1$, $w=0$, $w=1/3\pm 10^{-5}$, $w=2/3$ and $w=1-10^{-5}$.
Note that in the radiation dominant universe with $w=1/3$,
the action is given by
\begin{align}
S_2 
\big\vert_{w=1/3}
&=-\frac{m_{\rm eff}^2(2+\mu_{\pm})}{2\kappa_-^2} 
\int d\Omega 
\int d\eta dr (arH)^2 \left(\partial_{r}\pi\right)^2\,,
\end{align}
which does not describe the dynamics of $\pi$
 \footnote{ If we take into account the trace anomaly of quantum corrections, 
we find small deviation from $w=1/3$. For example, 
$w={1\over 3}-{5\over 18\pi^2}{g^4\over (4\pi)^2}
{(N_c+{5\over 4}N_f)({11\over 3}N_c-{2\over 3}N_f)
\over 2+{7\over 2}[N_cN_f/(N_c^2-1)]}$ 
for a plasma of the SU($N_c$) gauge theory with coupling $g$
and $N_f$ flavors \cite{trace anomaly}. }.
Thus the expansions we have adopted in our calculation is invalidated in this limit, and in order to correctly study the dynamics of the scalar graviton,
we must calculate the higher-order terms of $\pi$.

For $w\simeq 1/3$, the existence of the stable solution
is guaranteed for the parameter region such that
\begin{align}
\beta_2^2>\beta_3>\beta_2\,,
\quad &{\rm for}\;\; w=\frac{1}{3}-|\delta w|
\,,\\
\beta_3>\beta_2\,,
\quad &{\rm for}\;\; w=\frac{1}{3}+|\delta w|
\,,
\end{align}
with $|\delta w|\ll1$.
In such a parameter region, at least one of $\mu_{\pm}$
satisfies the stability condition \eqref{no_gg}
as well as our ansatz $\mu_{\pm}>-1$.

For $w=1$, $\mu_{\pm}$ are given by
\begin{align}
\mu_{\pm}= \frac{1-2\beta_2\pm |1-2\beta_2| }{2(2\beta_3-\beta_2)}\,.
\end{align}
Hence, one of $\mu_{\pm}$ becomes zero, which gives
 the homothetic solution.  

Since reducing the stability condition \eqref{no_gg} to the allowed parameter region for arbitrary values of $w$ is a nontrivial task due to the complicated dependence \eqref{mu_pm} of $\mu_\pm$ on the model parameters,
we analyze the stable region numerically. 
We conclude that the parameter region of
\begin{align}
\beta_2^2>\beta_3>\beta_2\,,
\label{stable_condition}
\end{align}
guarantees the existence of a stable branch for any values of $w$ except for $w=1/3$.
Even outside the region \eqref{stable_condition}, we obtain stable branches
for some values of $w$, but  the instability always appears
for near radiation dominant stage such that $w=1/3-|\delta w|$.

\begin{figure}[htb]
\centering
\includegraphics[width=8cm,angle=0,clip]{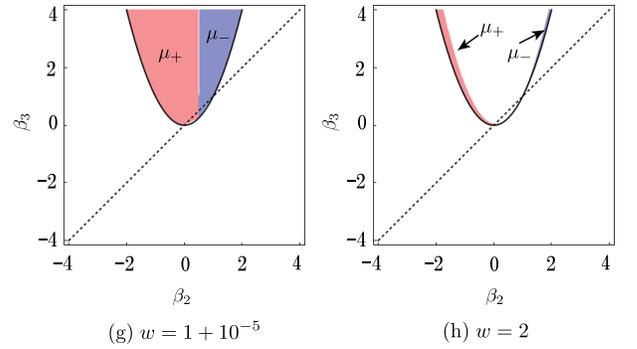}
\caption{The same figures as Fig. \ref{no_instability1}
for the cases of $w=1+10^{-5}$ and $w=2$.}
\label{no_instability2}
\end{figure}

In the stable region \eqref{stable_condition}, 
only one stable branch exists for any $w$.
When the Universe consists of the usual matter field with $w<1$, 
either $\mu_0=\mu_{+}$ or $\mu_{-}$ gives a stable solution
 in the parameter region \eqref{stable_condition}.
While if the Universe is
composed  effectively of a ``strange" matter field with $w> 1$,
we find $\mu_0=\mu_\pm$\,-\,branches are stable only in the 
region of $\beta_3>\beta_2^2$ (see Fig.~\ref{no_instability2}),
 and the stable region disappears
in the limit of $w=\infty$ as shown in Fig.~\ref{no_instability2}.
As a result, only the $\mu_0=0$\,-\,branch becomes stable.

Although the condition \eqref{stable_condition} depends on the proportional factor $K$,
we can derive the stability condition for the original coupling constants $\{b_i\}$.
Since we assume $m_{\rm eff}^2>0$ (i.e. $Kb_1+2K^2b_2+K^3b_3>0$),
the condition reduces to
\begin{align}
b_2^2-b_1b_3>0\,,
\quad
b_2<0\,.
\label{constraint_coupling}
\end{align}
Therefore,
if we choose the coupling constants
in the above parameter regions \eqref{constraint_coupling},
both ghost instability and 
gradient instability
are avoided because of the nonlinear interactions, 
and then the early stage of the Universe described 
approximately by GR FLRW spacetime becomes stable.



\section{Perturbations with matter effects}
\label{cosmology_with_nonlinear}
In this section we consider the adiabatic modes
retrieving matter perturbations 
and discuss the evolution history with the nonlinear effects of the St\"uckelberg fields.
For the stable branch discussed in the previous section,
the evolutions of the matter perturbations are approximated
by the adiabatic modes. The oscillation modes of matter fluctuations decay
 in time \cite{bigravity_dark_matter}. 
Hence we can decouple the adiabatic and oscillation modes in this case as well.
\\
~~
\subsection{Adiabatic mode with matter perturbations}
\label{adiabatic_with_matter}
Now we discuss the time evolution of the adiabatic modes with 
matter perturbations.
The matter energy densities are perturbed as
\begin{align}
\rho_g=\bar{\rho}_g(1+\delta_g)\,,
\quad
\rho_f=\bar{\rho}_f(1+\delta_f)\,.
\end{align}
We ignore pressure perturbations
and spatial velocities compared to 
the density perturbations, just for simplicity.

All equations are expanded up to the second order in $\epsilon$.
The Einstein equations for $g_{\mu\nu}$ and $f_{\mu\nu}$ give
\begin{widetext}
\begin{align}
2\Psi_g(\eta,r)&=a^2(\eta)r^2\left[m_g^2\left(\mu+\beta_2\mu^2+\frac{\beta_3}{3}\mu^3\right)
+\frac{1}{3}\kappa_g^2\bar{\rho}_g \tilde{\delta}_g \right]\,, 
\label{psi_g} \\
2r\frac{\partial \Phi_g}{\partial r}(\eta,r)&=a^2(\eta)r^2
\left[-m_g^2\left(\mu-\frac{\beta_3}{3}\mu^3\right)+
\frac{1}{3}\kappa_g^2 \bar{\rho}_g \tilde{\delta}_g \right]\,,
\label{phi_g}
\end{align}
and
\begin{align}
2\Psi_f(\eta_f,r_f)&=a^2(\eta_f)r_f^2\left[-\frac{m_f^2}{(1+\mu)^3}
\left( \mu+(1+\beta_2)\mu^2+\frac{1+\beta_2+\beta_3}{3}\mu^3\right)
+\frac{1}{3}K^2\kappa_f^2\bar{\rho}_f\tilde{\delta}_f\right]\,, 
\label{psi_f}  \\
2r_f\frac{\partial \Phi_f}{\partial r_f}(\eta_f,r_f)&=a^2(\eta_f)r_f^2\left[
\frac{m_f^2}{(1+\mu)^3}\left(\mu+2\mu^2+\frac{2+2\beta_2-\beta_3}{3}\mu^3\right)
+\frac{1}{3}K^2\kappa_f^2\bar{\rho}_f \tilde{\delta}_f\right]\,,
\label{phi_f}
\end{align}
respectively, where 
\begin{align}
\tilde{\delta}_g(\eta,r)&:=\frac{\int^{r}_0 4\pi \tilde{r}^2 \delta_g d\tilde{r}}
{\int^{r}_0 4\pi \tilde{r}^2 d\tilde{r}}\,, \quad
\tilde{\delta}_f(\eta_f,r_f):=\frac{\int^{r_f}_0 4\pi \tilde{r}^2 \delta_f d\tilde{r}}
{\int^{r_f}_0 4\pi \tilde{r}^2 d\tilde{r}}
\,,
\label{average_density}
\end{align}
are spatial averages of the density perturbations in the spheres with the radii $r$ and $r_f$,
respectively.
We define the mass parameters by
\begin{align}
m_g^2&:=\frac{m^2\kappa_g^2}{\kappa^2}(b_1K+2b_2K^2+b_3K^3)\,, \quad
m_f^2:=\frac{m^2\kappa_f^2}{K^2\kappa^2}(b_1K+2b_2K^2+b_3K^3)\,,
\end{align}
with which the effective graviton mass is expressed by $m_g^2$ and $m_f^2$ as
$m_{\rm eff}^2=m_g^2+m_f^2$.

Although the $f$-variables are given as the functions of $(\eta_f,r_f)$,
it is easy to find them as the functions of $(\eta,r)$
by use of the St\"uckelberg fields $\nu$ and $\mu$.
The variable $\nu$ is determined by \eqref{nu_eq}
even when the matter perturbations are included.

Substituting Eqs.~\eqref{psi_g}-\eqref{phi_f}
and \eqref{nu_eq} into the $r$ component of $\nabla_{\alpha}T^{[\gamma]\alpha}{}_{\beta}=0$,
we obtain an algebraic equation for $\mu$:
\begin{align}
&\mathcal{C}_{m^2}(\mu)+\mathcal{C}_{H^2}(\mu)+
\mathcal{C}_{\rm matter}(\mu)=0 \label{mu_eq}\,,
\end{align}
where we define each function as
\begin{align}
\mathcal{C}_{m^2}(\mu)&:=\mu \Big\{
m_g^2 (1+\mu)^2
        \left[ 9+18\beta_2\mu+(6\beta_2^2+4\beta_3)\mu^2-\beta_3^2\mu^4 \right]
\nonumber \\
&\qquad 
+m_f^2 \Big[ 9+18(1+\beta_2)\mu
       +(10+34\beta_2+4\beta_3+6\beta_2^2)\mu^2 \nonumber \\
&\qquad \qquad \quad
        +(2+14\beta_2+8\beta_3+12\beta_2^2)\mu^3  
        +(2\beta_2+2\beta_3+2\beta_2^2-\beta_3^2+4\beta_2\beta_3)\mu^4 \Big] 
\Big\}\,, 
\label{Fm2}\\
\mathcal{C}_{H^2}(\mu)
&:=-3H^2\mu(1+\mu)^2 \left\{ 3(1-w)
+2\left[1+(1-3w)\beta_2\right]\mu
+\left[ 2\beta_2-(1+3w)\beta_3 \right] \mu^2 \right\}\,,\\
\mathcal{C}_{\rm matter}(\mu) &:=
(1+\mu)^2 \left\{ \kappa_g^2\rho_g \tilde{\delta}_g(1-\beta_3 \mu^2)
-K^2\kappa_f^2\rho_f \tilde{\delta}_f (1+\mu)
\left[ 1+2\mu+(2\beta_2-\beta_3)\mu^2 \right]
\right\} \,.
\label{F_delta_matter}
\end{align}
\end{widetext}
Eq.~\eqref{mu_eq} reproduces the static result \eqref{mu_eq2}
in the limit of $H\rightarrow 0$.

An important difference from the case without matter perturbations
is that the variable $\mu$ depends also on the matter perturbations.
When the matter perturbations are not negligible, Eq.~\eqref{mu_eq_without_matter}
should be replaced with
Eq.~\eqref{mu_eq}, in which 
the typical value of the additional third term is evaluated by
\begin{align*}
\mathcal{C}_{\rm matter} \sim 
\kappa_g^2\rho_g\tilde{\delta}_g-K^2\kappa_f^2\rho_f \tilde{\delta}_f
\,.
\end{align*}

The metric perturbations are given by the GR results with
the corrections coming from the interaction term,
e.g., one of the perturbations of $g_{\mu\nu}$ is given by
\begin{align}
\Psi_g=\Psi_{\rm GR}+
a^2r^2m_{\rm eff}^2 \times \frac{\kappa_g^2}{\kappa_-^2}
\times \mathcal{O}(\mu)\,.
\label{gravity_with_mass}
\end{align}
When the second term is negligible compared to the first one, 
the metric perturbations are restored to the GR results.
Since the equations of motion of twin matters are
not modified from usual ones (e.g., see Eq. \eqref{c1}), 
the restoration of the metric perturbations
guarantees the dynamics of the matter
is also restored to the GR result.
Therefore we will discuss only the metric perturbations
which are determined by $\mu$ as in Eqs.~\eqref{psi_g}-\eqref{phi_f}.

\subsection{GR phase}
\label{sec_GR_phase}
We discuss a stage 
when the Hubble parameter is larger than the effective graviton mass
(i.e., $H^2\gg m_{\rm eff}^2$).
Because  $\mathcal{C}_{H^2}\gg \mathcal{C}_{m^2}$,
Eq.~\eqref{mu_eq} becomes
\begin{align}
\mathcal{C}_{H^2}+\mathcal{C}_{\rm matter}\approx 0\,. 
\label{mu_GR}
\end{align}
Since the second term is much smaller than the first term,
Eq.~\eqref{mu_GR} is schematically solved as
\begin{align}
\mu=\mu_0+\mathcal{O}(\tilde{\delta}_g,\tilde{\delta}_f)\,,
\end{align}
where $|\tilde{\delta}_g|,|\tilde{\delta}_f|\ll 1$.
As discussed in Sec.~\ref{perturbation_around_FLRW},
the stable branch is found with $\mu_0=\mu_{\pm}$ for $w<1$,
or with $\mu_0=0$ for $w>1$.

First, we consider the case of $w<1$.
The stable branch is given by one of $\mu_{\pm}$,
and thus \eqref{mu_GR} is also solved as $\mu \approx \mu_{\pm}$.
The gravitational sector is restored to the one in GR,
when 
\begin{align}
\frac{\kappa_g^2}{\kappa_-^2}\times \frac{m_{\rm eff}^2}{H^2}\ll \tilde{\delta}_g
\,,
\end{align}
i.e., if the correction terms from the graviton mass are negligible compared to the  GR terms
 in Eqs.~\eqref{psi_g}-\eqref{phi_f}.
As a result, in the early stage of the Universe,
the metric perturbations are restored to the GR results.

Next, we consider the case of $w>1$,
in which the stable branch is given by $\mu_0=0$.
In this case, the solution is given by,
assuming $\vert \mu \vert \ll 1$,
\begin{align}
\mu\approx \frac{\tilde{\delta}_g
-\tilde{\delta}_f}
{3(1-w)}\,,
\label{bigravity_mu_linear}
\end{align}
where we have used the background equations.
One of the metric perturbations is described as
\begin{align}
2\Psi_g=a^2r^2H^2\tilde{\delta}_g
+a^2r^2m_{\rm eff}^2\frac{\kappa_g^2}{\kappa_-^2}\frac{\tilde{\delta}_g
-\tilde{\delta}_f}
{3(1-w)}\,.
\end{align}
Since the second term is negligible compared to the first term
in the case of $H\gg m_{\rm eff}$,
the metric perturbations are again restored to the GR results for $w>1$.

Hence both cases show the GR limit in the early stage of the Universe
($H\gg m_{\rm eff}$).
We shall call this stage the GR phase.

\subsection{Bigravity phase}
\label{bigravity_phase}
Secondly, we discuss the stage 
when the Hubble parameter is smaller than the effective graviton mass 
(i.e., $H^2\ll m_{\rm eff}^2$).
In this stage, we find $\mathcal{C}_{m^2}\gg\mathcal{C}_{H^2}$.
Hence, for the matter of our interest, Eq.~\eqref{mu_eq} reduces to
\begin{align}
\mathcal{C}_{m^2}+\mathcal{C}_{\rm matter}\approx 0\,. 
\label{mu_bigravity}
\end{align}
We denote the roots of $\mathcal{C}_{m^2}(\mu)=0$ by $\mu_{\infty}$,
which are found to be zero and some constants of order unity.
Similarly to the previous subsection,
Eq.~\eqref{mu_bigravity} is solved as
\begin{align}
\mu=\mu_{\infty}+\frac{H^2}{m_{\rm eff}^2}
\times \mathcal{O}(\tilde{\delta}_g,\tilde{\delta}_f)\,.
\end{align}

Since Eq.~\eqref{mu_bigravity} is a polynomial equation of degree seven
 for $\mu$,
there are seven solutions for $\mu_\infty$. 
We classify the solutions of Eq. (\ref{mu_bigravity}) into two types:
the linear branch and non-linear branches.
Note that a branch here denotes one 
with $\mu=\mu_\infty$ 
in the limit of $H/m_{\rm eff}\rightarrow 0$.

The linear branch is realized by choosing $\mu_{\infty} = 0$.
Eq. \eqref{mu_bigravity}  gives the value of $\mu$ as,
assuming $\vert \mu \vert \ll 1$,
\begin{align}
\mu\approx -\frac{
\kappa_g^2\bar{\rho}_g\tilde{\delta}_g
-K^2\kappa_f^2\bar{\rho}_f\tilde{\delta}_f}
{9m_{\rm eff}^2}\,.
\end{align}
Substituting this solution into the expression of the gravitational force
\eqref{phi_g},
we find
\begin{align}
\Phi'_g \approx 
\frac{a^2r}{6}\left[
\left(1+\frac{m_g^2}{3m_{\rm eff}^2}\right)\kappa_g^2\tilde{\rho}_g 
-\frac{m_g^2}{3m_{\rm eff}^2}K^2\kappa_f^2\tilde{\rho}_f\right] \; .
\end{align}
Hence the gravitational force in the $g$-sector  is produced by 
the $f$-matter as well as the $g$-matter.
We have done a detailed study including stability
about this linear branch
in our previous paper \cite{bigravity_dark_matter}.

The nonlinear branches are obtained by choosing $\mu_{\infty} \sim \mathcal{O}(1)$.
In this case, since $|\mu_{\infty}| \gg |\tilde{\delta}_g|, |\tilde{\delta}_f|$,
 the solutions are found to be 
\begin{align}
\mu \approx \mu_{\infty}\,,
\end{align}
giving
the same as those without matter perturbations.
For these nonlinear branches, 
the metric perturbations include the correction terms of 
the graviton masses
as given  in Eqs.~\eqref{psi_g}-\eqref{phi_f}.
Hence, for these branches, 
there is a non-negligible inhomogeneity at large scale
and the gravitational behaviour deviate largely from GR's one
beyond the Vainshtein radius
as shown in Appendix \ref{app_late_stability}.

We shall study the stability around the nonlinear branches as well as the linear branch
in the bigravity phase ($m_{\rm eff}^2 \gg H^2$).
In the limit $H\rightarrow 0$ or $H^2 \rightarrow \Lambda_g/3$,
we find the adiabatic solutions turn to be the static ones
given in Appendix \ref{flat_Vainshtein}.
In the sub-horizon scale, the cosmological coordinates $(\eta,r)$ are related 
to the static coordinate $(T,R)$ as
\begin{align}
T=t+\frac{1}{2}a^2r^2\sqrt{\frac{\Lambda_g}{3}}+\cdots \,,\quad 
R=ar\,,
\label{coordinate_trans}
\end{align} 
where $\displaystyle{t=\int ad\eta}$ is the cosmic time.
On the cosmological coordinate $(\eta,r)$, the St\"uckelberg variable $\mu$ is defined by
\begin{align}
\mu:=\frac{r_f-r}{r}\,,
\end{align}
while on the static coordinate $(T,R)$, $\hat{\mu}$ is defined by
\begin{align}
\hat{\mu}:=\frac{R_f-R}{R}\,.
\end{align}
Because of 
\begin{align}
\mu=\hat{\mu}\,,
\end{align}
for Eq.~(\ref{coordinate_trans}), 
we find  Eq.~\eqref{mu_eq} corresponds to Eq.~\eqref{mu_eq2}
even in the limit $H^2 \rightarrow \Lambda_g/3$.

Therefore, as a lowest-order approximation,
we regard the adiabatic solutions as the static ones
and we can apply the stability analysis in Appendix \ref{app_late_stability}
to the present case.
We note that the asymptotically homothetic branch of Eq.~\eqref{mu_eq2} 
corresponds to the linear branch in the bigravity phase,
while non-asymptotically homothetic branches correspond to 
nonlinear branches in the bigravity phase.
It turns out that  the linear branch is always stable.
Furthermore, we find 
that one of the nonlinear branch is stable for some coupling constants.

These arguments are essentially unchanged even when 
a cosmological constant 
with the condition of $\Lambda_g \lsim  3m_{\rm eff}^2/2 $ is introduced.
We give the results with such a cosmological constant 
for the gravitational behaviour 
and the Vainshtein screening
in Appendix \ref{flat_Vainshtein1},
and those for the stability in Appendix \ref{app_late_stability}.

Since the gravitational behaviours are modified from
the ones in GR
due to the existence of the fifth force,
mediated by the scalar mode of graviton,
for both branches,
we call this stage the bigravity phase.



\section{Transition from GR to bigravity}
\label{Sec_transition}
As we have shown in the previous section, the Universe in the bigravity theory 
has some stable branches in the both  limits of $H\gg m_{\rm eff}$ and of 
$H\ll m_{\rm eff}$, which correspond to the early   and late stages of the Universe,
respectively.  More precisely,
in the early stage of the Universe ($H\gg m_{\rm eff}$), 
the $\mu_0=0$\,-\,branch
is unstable, while  $\mu_0=\mu_\pm $\,-\,branches 
can give the stable GR phase depending on 
the coupling constants.
On the other hand, 
in the late stage of the Universe  ($H\ll m_{\rm eff}$), 
the branch with $|\mu|  \ll 1$,  
 which provide us the bigravity phase, is stable for 
appropriate coupling constants.
In addition, some branches with $\mu_\infty\sim O(1)$
  (the bigravity  phase) are also stable for some coupling constants.

The question is whether any two stable branches with different 
limits can connect under our adiabatic approximation or not. 
We shall discuss this possibility in this section.

Since our Universe is homogeneous at large scale,
the St\"uckelberg variables should reach to the linear bigravity phase
as the Universe expands.
Hence our Universe must start from the GR phase and 
transit to the bigravity phase.
Then the Universe must pass through the period of $H  \sim  m_{\rm eff}$, 
where the behaviour of $\mu$ becomes unclear.

One unknown in this period is the transition time scale, if transition occurs.
If the transition time scale is given by the Hubble time scale
 and the adiabatic mode is an attractor even in the period of $H\sim m_{\rm eff}$,
we can discuss the transition by considering only the adiabatic modes.
However, 
if two stable adiabatic solutions in GR and in bigravity phases 
are  discontinuous, 
the adiabatic approximation breaks down, and then 
the transition time scale may be faster than $H^{-1}$ even if such a transition 
is possible. 
As a result, 
the assumption \eqref{adiabatic_assumption} is no longer valid,
and a full analysis without approximation will be required.

Here, we only speculate some possible transitions based on the adiabatic approximation
with the assumption \eqref{adiabatic_assumption}.
When the amplitudes of the density perturbations are given,
one can obtain the St\"uckelberg modulation $\mu$ in terms of $H$ and $\tilde \delta_{g/f}$ by solving the algebraic equation \eqref{mu_eq} together with the use of \eqref{Friedmann_background} and \eqref{eq_matter}.
Since $H$ decreases in time, we find the evolution of $\mu$ 
without solving the equations of motion, under the assumption that the solutions to \eqref{mu_eq} at different moments are continuously connected.
Here we also assume that the density perturbations are constant in time;
although they may evolve in time,
the qualitative behaviour does not change much.

\begin{figure}[tbp]
\begin{center}
\includegraphics[width=7.3cm,angle=0,clip]{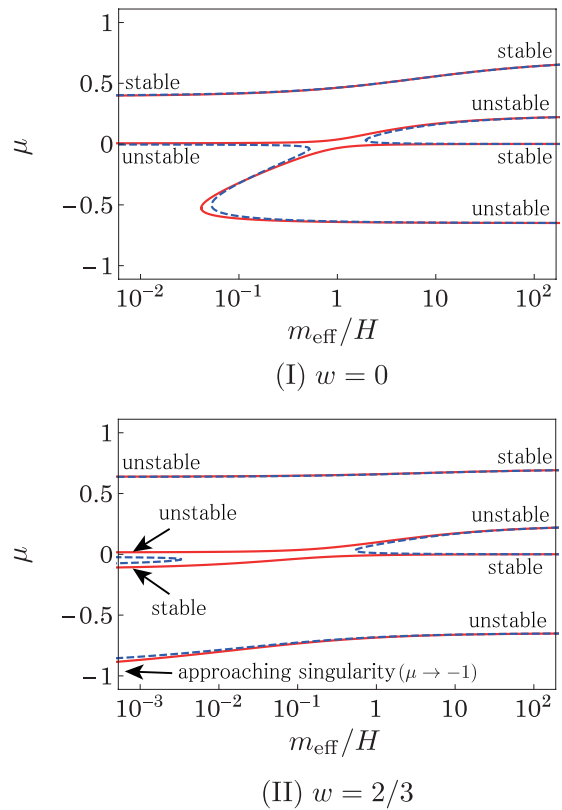}
\caption{Case A: There exists the stable branch with 
non-linear $\mu_\infty (\sim O(1))$.
We plot the roots of Eq. \eqref{mu_eq}
for $\tilde{\delta}_g=10^{-2}, \tilde{\delta}_f=10^{-3}$ 
(red curves)
and for $\tilde{\delta}_g=10^{-3}, \tilde{\delta}_f=10^{-2}$ 
(blue dashed curves).
We set $\beta_2=-3,\beta_3=3, m_g^2=m_f^2$.
The branch with $\mu_0 \simeq 0.40$ for   $w=0$ 
and $\mu_0 \simeq -0.10$ for  $w=2/3$
are stable in the early stage of the Universe ($m_{\rm eff}/H\ll 1$).
In the late stage ($m_{\rm eff}/H\gg 1$), 
the branches with $\mu_0 \simeq 0$ and $0.70$ are stable.
The Universe may evolve from the stable $\mu_0$\,-\,branch 
to the stable $\mu_\infty$\,-\,branch.
For example, for $w=0$, $\mu$ changes from 
$\mu_0=0.4$ (GR phase) to $\mu_\infty =0.7$ (nonlinear bigravity phase),
while  for $w=2/3$, it does 
 from $\mu_0 =-0.1$ (GR phase) to $\mu_\infty =0$ (linear bigravity phase).
}
\label{transition_fig}
\end{center}
\end{figure}

\begin{figure}[tbp]
\begin{center}
\includegraphics[width=7cm,angle=0,clip]{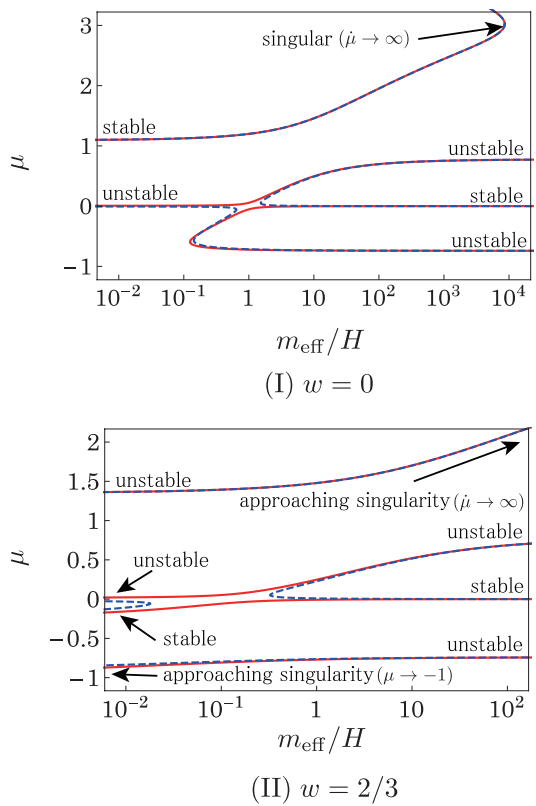}
\caption{Case B: There exists only one stable branch with 
$\mu_\infty=0$.
We plot the roots of Eq.~\eqref{mu_eq}
for $\tilde{\delta}_g=10^{-2}, \tilde{\delta}_f=10^{-3}$ 
(red curves)
and $\tilde{\delta}_g=10^{-3}, \tilde{\delta}_f=10^{-2}$ 
(blue dashed curves).
We set $\beta_2=-1, \beta_3=1/2, m_g^2=m_f^2$.
The stable branches
for  $m_{\rm eff}/H\ll 1$ are given by
$\mu_0 \simeq 1.1$ for $w=0$,
and $\mu_0 \simeq -0.21$ for $w=2/3$.
For $m_{\rm eff}/H\gg 1$, 
only $\mu_\infty \simeq 0$ is stable.
The Universe with  $w=2/3$ may evolve from 
$\mu_0=-0.21$  (GR phase) to $\mu_\infty =0$ (linear bigravity phase),
but  for $w=0$, there is no stable adiabatic solution.
}
\label{transition_fig2}
\end{center}
\end{figure}

We show several examples in Figs.~\ref{transition_fig} and \ref{transition_fig2}.
We have explored a wide  range of the  parameters with 
the stability condition \eqref{stable_condition}.
The stability analysis 
for the late stage of the Universe ($m_{\rm eff}/H\gg 1$)
is given in Appendix \ref{app_late_stability}.

As shown in Figs.~\ref{transition_fig} and \ref{transition_fig2},
two stable branches in the limits of  $m_{\rm eff}/H\ll 1$
and of $m_{\rm eff}/H\gg 1$
can be continuous or discontinuous depending on 
the equation-of-state parameter $w$ and the density perturbations.
For example, 
as shown in Fig.~\ref{transition_fig} (II), 
for $w=2/3$, we find one continuous curve from  the early stage to the late stage 
of the Universe,
exhibiting the branch that is stable in both limits of  $m_{\rm eff}/H\ll 1$
and of $m_{\rm eff}/H\gg 1$, if $\tilde{\delta}_g(=10^{-2})>\tilde{\delta}_f(=10^{-3})$.
On the other hand, if $\tilde{\delta}_g(=10^{-3})<\tilde{\delta}_f(=10^{-2})$,
two (one stable and another unstable) continuous curves are splitted
into two discontinuous curves as found  in Fig.~\ref{transition_fig} (II).
There are points at which
two real roots of Eq.~\eqref{mu_eq} degenerate
and beyond which they become imaginary.
At such points, the time derivative of $\mu$ ($d\mu/dH^{-1}$) diverges,
but $\mu$ itself is finite.
At such a singular point, we argue that 
the energy density of the scalar graviton diverges.
This is because the variable $\mu$ is basically related to 
the adiabatic mode of the scalar graviton 
$\pi_0$ as $\mu=a^{-2}\pi_0'/r$.
Similarly to the Vainshtein screening mechanism, 
the derivative of $r\mu$ ($ \sim \partial \partial \pi_0$)
gives a typical energy density of the scalar graviton.
So when $\partial\mu\rightarrow \pm \infty$ (or  
$\partial\partial\pi \rightarrow \infty$), 
a singular behaviour of the scalar graviton may appear.

However, near such a singular point, our assumption \eqref{adiabatic_assumption}
is no longer valid.
This singular behaviour may simply be an artifact of our adiabatic assumption, and
if we analyze the full nonlinear dynamics without the adiabatic approximation,
this singularity may not appear.
Hence,  when we find a discontinuity in the solution $\mu (H)$, 
what happens in the transition period is still an open problem.
In what follows, we just discuss the adiabatic 
solutions given by the continuous curves.

We classify our results  into two cases 
by whether the stable nonlinear branch 
in the late stage of the Universe exists (Case A) or not (Case B).
Some examples of Case A are given in Fig.~\ref{transition_fig},
while those of  
Case B are in Fig.~\ref{transition_fig2}.
We shall discuss them in due order.

\noindent
{\bf (1)  Case A}:\\
For $w<1/3$ (Fig.~\ref{transition_fig} (I)),
there exists a stable $\mu_0$\,-\,branch in the early stage of the Universe
($H\gg m_{\rm eff}$), and nonlinear stable $\mu_\infty$\,-\,branch 
in the late stage ($H\ll m_{\rm eff}$).
Hence we expect that the Universe will evolve from GR phase 
to non-linear bigravity phase.
The GR phase in the early stage
transits to the inhomogeneous universe in the late stage.

Our present stability analysis, however,
is valid only for both limits of  $H\gg m_{\rm eff}$ and 
of $H\ll m_{\rm eff}$.
Therefore, it is not immediate to decide whether  
our adiabatic solution is still stable
at $H\sim m_{\rm eff} $.
If the adiabatic solution becomes unstable near the period of $H\sim m_{\rm eff} $,
the adiabatic mode does not evolve independently of the oscillation mode.
Then the transition time scale becomes faster than $H^{-1}$.
As a result, the GR phase in the early stage 
might connect to the linear bigravity phase in the late stage.

On the other hand, in the case of  $w>1/3$, 
only for the case of $\tilde{\delta}_g>\tilde{\delta}_f$, 
there exists only one stable branch from $\mu_0$ to $\mu_\infty$.
It may describe that the GR phase in the early stage of the Universe 
transits to the linear bigravity phase in the late stage.
If  $\tilde{\delta}_g<\tilde{\delta}_f$,
 two stable branches are disconnected.
The assumption \eqref{adiabatic_assumption}
may no longer be valid at the singular points.

\noindent
{\bf (2) Case B}:\\
For the case of $w>1/3$, we find
similar behaviours to Case A.
There exists one stable branch from 
the GR phase in the early stage of the Universe 
transits to the linear bigravity phase in the late stage,
if $\tilde{\delta}_g>\tilde{\delta}_f$.
 If  $\tilde{\delta}_g<\tilde{\delta}_f$, however, 
 two stable branches are disconnected.

For $w<1/3$,  the stable GR branch evolves into a singularity.
No other stable branch exists.
Since the assumption \eqref{adiabatic_assumption}
may no longer be valid, 
the transition from the GR phase 
to the linear bigravity phase might be possible,
 if the stable $\mu_\infty=0$\,-\,branch  is
a global attractor.

From the above analysis, 
we conclude that for $w>1/3$,
there exists a stable adiabatic solution,
which describes that the Universe evolves from 
a GR phase in the early stage to a linear  bigravity 
phase in the late stage.
For more realistic equation-of-state parameter $w<1/3$,
however, 
if the transition from the GR phase 
to the linear bigravity phase occurs, 
the transition time scale is likely to be faster than $H^{-1}$ 
and then the adiabatic condition must no longer be valid
\footnote{One may argue the possibility that 
the transition might occur with a Hubble time scale
by taking into account the evolutions of the density perturbations.
However this is unlikely because
the qualitative behaviours of Figs.~\ref{transition_fig} and \ref{transition_fig2}
do not depend on the amplitudes of $\tilde{\delta}_g$ and $\tilde{\delta}_f$
but rather on their ratio 
$\tilde{\delta}_g/\tilde{\delta}_f$.}.
In order to conclusively analyze the full evolution history of the Universe 
for natural  equation-of-state parameter $w$, 
the analysis beyond the adiabatic approximation is required,
and a numerical treatment should be expected, which is the work in progress.

\section{Concluding Remarks}
\label{summary}
Since the homogeneous and isotropic universe is consistent with 
cosmological observations, the existence of 
exponential instability around the FLRW spacetime
would ruin the viability of such cosmological solutions,
unless the initial condition is fine-tuned.
It has been known that in the bigravity theory, while the FLRW background solutions are stable against small perturbations at the late stage of the Universe ($H \ll m_{\rm eff}$), the linear perturbations suffer either a ghost or a gradient instability during the early epoch ($H\gg m_{\rm eff}$).
However, the unstable mode arises from the St\"uckelberg fields
which determine the relations between the two coordinates of the physical and fiducial sectors in the bigravity.
The instability implies that
the nonlinear interactions of the St\"uckelberg fields may be significant.
However, the nonlinearities of the St\"uckelberg fields
do not directly conclude that
the spacetime perturbations around
the homogeneous and isotropic spacetime are large.
For this reason,
we study small metric perturbations around the flat FLRW background 
retaining nonlinearities of the St\"uckelberg fields
and discuss whether or not the viable cosmological solution exists even in 
$H\gg m_{\rm eff}$.

In order to comfort the difficulty in
the full nonlinear analysis,
we consider the following simplifications:
\begin{enumerate}
\item
We assume both background spacetimes are almost isotropic and homogeneous.
This background is regarded as
a ground state of the massive graviton.
\item
Since the scalar graviton mode is essential to the Higuchi ghost and
the gradient instability,  
we restrict our analysis to spherically symmetric configurations, in which the degrees of freedom of the tensor and the vector gravitons do not exist,
while the scalar graviton propagates.
\item
We decompose the perturbation variables into adiabatic modes and oscillation modes, 
and assume that the adiabatic modes evolve independently of the oscillation modes.
The later assumption is valid only if the oscillation modes are stable.
\item
We only consider the sub-horizon scale and assume that the adiabatic modes satisfy the condition \eqref{adiabatic_assumption}.
Under these assumptions, 
we  find that
the result in the adiabatic modes reproduce the static result \eqref{mu_eq2}
in the limit $H\rightarrow 0$.
\end{enumerate}

Then,
we show that the Higuchi type ghost instability and the gradient instability
can be removed by the nonlinear effects of the scalar graviton.
In the early stage of the Universe $(H\gg m_{\rm eff})$,
we find a few sets of cosmological solutions \eqref{cos_sol},
in which there are two types;
one is given by the homothetic spacetimes, and 
other branches are given by the approximately homogeneous and isotropic spacetimes
on the different coordinates.
In the  former homothetic solutions, the instability cannot be avoided
when the Universe consists of ordinary matter components.
On the other hand, 
the solutions in the other branches
can avoid both instabilities 
if the coupling constants satisfy the simple inequalities \eqref{constraint_coupling}.
For any values of the equation of state, there exists at least one stable branch, in which the Universe evolves in the same way as in GR.
The result suggests that
the bigravity has the healthy massless limit,
and cosmology based on the bigravity 
is still viable even in the early stage of the Universe.

Hence we conjecture the following cosmic scenario:
First, the cosmic growth history is restored to the GR result 
in the early stage  $(H\gg m_{\rm eff})$,
in which the St\"uckelberg fields are non-linear.
After a transition around $H \sim m_{\rm eff}$, 
the Universe reaches the bigravity phase,
in which the modification from GR appears.
In the bigravity phase, 
the $\mu_\infty=0$\,-\,branch  is stable.
There is also another stable branch for some combinations of the coupling constants,
showing somewhat peculiar gravitational behaviours.

The transition from the GR phase to the bigravity phase
is, however, still an open problem.
One important key point is the transition time scale.
As discussed in Sec.~\ref{Sec_transition}, 
if the GR phase connects to the linear ($\mu_\infty = 0$) bigravity phase
with natural equation-of-state parameter ($w<1/3$),
the transition time scale should be faster than $H^{-1}$.
To analyze this transition phase, a numerical treatment should be required.

As we have shown, the GR cosmology is recovered in the early stage of the Universe, 
which we call the cosmological Vainshtein mechanism.
Since this is different from the Vainshtein screening mechanism 
discussed in some scalar-tensor type gravity theories such as 
the Galileon model, here we show how they are different.

In the present bigravity model, when we take a decoupling limit,
we find the effective action of a scalar field $\phi$, which corresponds
to the massive scalar graviton, as
\begin{eqnarray}
S_{\rm eff}&=&\int d^4x \sqrt{-g}\Big[
 {\cal L}_2^{(\phi)}+{\cal L}_{\rm NL}^{(\phi)}
\nn
&+&
\frac{R^{\mu\nu}}{m_{\rm eff}^2}{\cal L}_{2, \mu\nu}^{(\phi)}+
\frac{R^{\mu\nu\rho\sigma}}{m_{\rm eff}^2}{\cal L}_{{\rm NL}\mu\nu\rho\sigma}^{(\phi)}
+\cdots\Big]
\,,~~~
\end{eqnarray}
where 
\begin{eqnarray}
&&{\cal L}_2^{(\phi)}=-{3\over 4}(\partial \phi)^2
\nn
&&
{\cal L}_{\rm NL}^{(\phi)}={c_{\rm NL} \over \Lambda^3}(\partial \phi)^2\Box \phi
\nn
&&
{\cal L}_{2, \mu\nu}^{(\phi)}={1 \over 2}\partial_\mu \phi\partial_\nu \phi
\nn
&&
{\cal L}_{{\rm NL}\mu\nu\rho\sigma}^{(\phi)}={\tilde{c}_{\rm NL}\over \Lambda^3}
\partial_\mu \phi\partial_\rho\phi\,\partial_\nu \partial_\sigma \phi
\,,
\end{eqnarray}
where $c_{\rm NL}, \tilde{c}_{\rm NL}$ are some dimensionless constants, and $\Lambda$ is 
a typical scale for a strong coupling state.
We have  dropped higher interaction terms.
In the conventional Vainshtein screening, the first two terms 
play the essential roles. When the energy scale of the scalar field is 
larger than the string coupling scale $\Lambda$ below the Vainshtein radius, 
the degree of freedom of the scalar field is killed by the 
second non-linear term ${\cal L}_{\rm NL}^{(\phi)}$, and then
GR is recovered. On the other hand, 
in the present cosmological Vainshtein screening,
the third and fourth terms play important roles. 
We find the GR solution in the early stage of the Universe as follows:
Comparing the first and third terms, 
when the typical scale of the curvature is larger than the graviton mass, 
we find the GR universe as a consequence. 
However, to stabilize this GR universe, we have to take into account the fourth 
non-linear term ${\cal L}_{{\rm NL}\mu\nu\rho\sigma}^{(\phi)}$ as well. 
Hence in the cosmological Vainshtein screening, the curvature terms are essential,
which is different from the conventional Vainshtein mechanism.

Our Vainshtein mechanism is in some respects similar to the ghost condensate \cite{ghost_condensate}.
In the ghost condensate, even if a scalar field has a ghost instability
at linear level,
non-zero time derivatives of the background expectation value
can stabilize the fluctuation of the scalar field.
Indeed, in our system,
assuming that the scalar graviton $\phi$ can be decomposed into 
the adiabatic mode $\pi_0$ and the oscillation mode $\pi$ as $\phi=\pi_0+\pi$,
the non-zero $\pi_0$ stabilizes 
the fluctuation $\pi$.

However there are some differences between the ghost condensate 
and the stabilization in the bigravity. 
First, we note that
the non-zero $\pi_0$ can stabilize not only the ghost instability 
but also the gradient instability
in the bigravity.
Furthermore, 
the scalar graviton is stabilized by
the higher order non-linear effects of the spatial derivative of $\pi_0$ 
as distinct from the time derivative in the ghost condensate.
This is because $\mu_0$ is related to $\pi_0$ as $\mu_0=a^{-2} \partial_r \pi_0/r$,
and the higher-order terms of $\mu_0$ are essential to the stabilization. 
Although the scalar graviton has inhomogeneity,
the spacetime is still almost 
homogeneous due to the existence of the screening mechanism.

In order to check complete viability of the cosmological model in the bigravity,
we must consider beyond spherically symmetric perturbations
as well as beyond our simple ansatz on the background.
The kinetic term \eqref{kinetic_curved} indicates
that the ghost or the gradient instability may also exist
even for non-spherical small perturbations
around the FLRW background.
Thus extended studies are required.

The stabilities of the vector graviton and the tensor graviton
are also non-trivial questions.
The gravitational field equations are restored to the Einstein equations in GR
when $H\gg m_{\rm eff}$.
Hence, the tensor mode should be
reduced to the GR one,
and then we expect that no instability may appear, just as in GR.
We also expect no exponential instability in the vector graviton,
because the kinetic term for the vector graviton is 
not modified even for a curved background at linear level
as shown in the equation \eqref{FP_with_St}.
The complete stability analysis,
however, should be done to confirm our speculation, which we leave for future work.

\section*{Acknowledgments}
We would like to thank Antonio De Felice, Lavinia Heisenberg, Shinji Mukohyama and Ryo Saito for useful discussions and comments.
 This work was supported in part by Grants-in-Aid from the 
Scientific Research Fund of the Japan Society for the Promotion of Science 
(Nos. 25400276 and 15J05540).

\newpage
~

\appendix

\section{Vainshtein screening  in a spherically-symmetric static spacetime}
\label{flat_Vainshtein}
In this Appendix, we consider the Vainshtein screening mechanism in 
a spherically-symmetric static spacetime. 
While the Vainshtein screening in bigravity for a spherically-symmetric static spacetime
was discussed in \cite{Vainshtein},
the effects due to the existence of the $f$-matter and cosmological constant have not been taken into account. We would like to include these effects in our present analysis.

\subsection{Basic equations}
Spherically-symmetric static spacetimes in the bigravity theory 
are assumed to take the two metrics of the form
\begin{align}
ds_g^2&=-e^{2\hat{\Phi}_g}dT_g^2+e^{2\hat{\Psi}_g}dR_g^2+R_g^2d\Omega^2\,, \\
ds_f^2&=K^2\left[-e^{2\hat{\Phi}_f}dT_f^2+e^{2\hat{\Psi}_f}dR_f^2+R_f^2d\Omega^2\right]\,.
\end{align}
In what follows, we fix the gauge as
\begin{align}
T_g&=T\,,\quad R_g=R\,.
\end{align}
Since the spacetimes are static, a non-diagonal component of $f_{\mu\nu}$
on the coordinates $(T,R)$ should vanish 
to find a non-trivial solution \cite{Vainshtein, massive BH}.
Hence $T_f$ must be proportional to $T$.
In addition, we expect two spacetimes are asymptotically homothetic at large distances
($R\rightarrow \infty$), where
the proportional factor $K$ is given by 
Eq.~(\ref{eq_K2}). 
So we set 
\begin{align}
T_f&=T
\,,\quad
R_f=R \left[1+\hat{\mu}(R) \right]\; ,
\end{align}
with the property $\hat \mu (R) \rightarrow 0$ in the limit $R \rightarrow \infty$.

As in \cite{Vainshtein}, we consider 
the region inside the Compton wavelength of the massive graviton and 
weak gravitational fields such that
the metrics and their spatial derivatives satisfy the following conditions:
\begin{align}
|\hat{\Phi}_g|,|\hat{\Psi}_g|,|\hat{\Phi}_f|,|\hat{\Psi}_f| &\ll 1\,,
\label{weak_gravity1}
\\
|R\hat{\Phi}_g'|,|R\hat{\Psi}_g'|,|R\hat{\Phi}_f'|,|R\hat{\Psi}_f'| &\ll 1\,,
\label{weak_gravity2}
\end{align}
where 
a prime denotes the derivative with respect to $R$,
which is used only in this Appendix.
Although  the variable $\hat{\mu} $ is not necessarily small,
we assume that 
$\hat{\mu} $ should  not be so large, satisfying 
\begin{align}
| \hat\mu \hat{\Phi}_g|\ll 1\,, \quad | R \hat\mu' \hat{\Phi}_g| \ll 1\,,
~~\cdots \,.
\label{assumption_flat_V}
\end{align}

From the Einstein equations, we obtain
\begin{widetext}
\begin{align}
\frac{\hat{\Psi}_g}{R^2}&=
\frac{1}{6}\kappa_g^2\tilde{\rho}_g +\frac{1}{6}\Lambda_g
+\frac{m_g^2}{2}\left(\hat{\mu}+\beta_2\hat{\mu}^2+\frac{\beta_3}{3}\hat{\mu}^3\right) 
\label{flat_psig}\,,
\\
\frac{\hat{\Psi}_f}{R_f^2}&=
\frac{1}{6}K^2\kappa_f^2\tilde{\rho}_f +\frac{1}{6}\Lambda_g
-\frac{m_f^2}{2(1+\hat{\mu})^3}
\left[ \hat{\mu}+(1+\beta_2)\hat{\mu}^2+\frac{1+\beta_2+\beta_3}{3}\hat{\mu}^3\right] \,,
\\
\frac{1}{R}\frac{d\hat{\Phi}_g}{dR}&=
\frac{1}{6}\kappa_g^2\tilde{\rho}_g
-\frac{1}{3}\Lambda_g
-\frac{m_g^2}{2}\left(\hat{\mu}-\frac{\beta_3}{3}\hat{\mu}^3\right)
\,,
\label{flat_phig}\\
\frac{1}{R_f}\frac{d\hat{\Phi}_f}{dR_f}&=
\frac{1}{6}K^2\kappa_f^2\tilde{\rho}_f
-\frac{1}{3}\Lambda_g
+\frac{m_f^2}{2(1+\hat{\mu})^3}
\left(\hat{\mu}+2\hat{\mu}^2+\frac{2+2\beta_2-\beta_3}{3}\hat{\mu}^3\right)\,,
\label{flat_phif}
\end{align}
where we introduce the mean densities in the spheres with the radii $R$ and $R_f$ by 
\begin{align}
\tilde{\rho}_g(R)
={\displaystyle{\int^{R}_0 4\pi \tilde{R}^2 \rho_g(\tilde{R}) d\tilde{R}}\over 
\displaystyle{\int^{R}_0 4\pi \tilde{R}^2 d\tilde{R}}}\,,
 \quad
\tilde{\rho}_f(R_f)
={\displaystyle{\int^{R_f}_0 4\pi \tilde{R}^2 \rho_f d\tilde{R}}\over 
\displaystyle{\int^{R_f}_0 4\pi \tilde{R}^2 d\tilde{R}}}\,.
\end{align}
with $\rho_g=-T_g^{[{\rm m}]\,T}{}_T$ and 
$\rho_f=-{\cal T}_f^{[{\rm m}]\,T_f}{}_{T_f}$, respectively.
We ignore the pressures of twin matters, for simplicity.

Substituting them into $\nablag{}_{\mu}T^{[\gamma]\mu}{}_{\nu}=0$, 
in the weak field limit, 
we find an algebraic equation for $\hat \mu$ as
\begin{align}
&\mathcal{C}_{m^2}(\hat{\mu})+\mathcal{C}_{\Lambda}(\hat{\mu}) 
+\mathcal{C}_{\rm matter}(\hat{\mu})=0 \label{mu_eq2}
\end{align}
where $\mathcal{C}_{m^2}$  and $\mathcal{C}_{\rm matter}$ are
 given by Eqs.~\eqref{Fm2}
and \eqref{F_delta_matter},
and 
\begin{align}
\mathcal{C}_{\Lambda}(\hat{\mu})&:=
-\Lambda_g\hat{\mu}(1+\hat{\mu})^2
\Bigl[6+(2+8\beta_2)\hat{\mu}+2(\beta_2+\beta_3)\hat{\mu}^2\Bigl]  
\,.
\end{align}

Eq.~\eqref{mu_eq2} is also found from 
 Eq.~\eqref{mu_eq} when we take the limit of $H^2 \rightarrow \Lambda_g/3$.
Hence we conclude that 
the cosmological adiabatic solution discussed in the text 
 corresponds to the present static solution in the ``static" 
limit of $H^2 \rightarrow \Lambda_g/3$.

\subsection{Vainshtein screening}
\label{flat_Vainshtein1}
When we assume $|\hat{\mu}| \ll 1$, 
Eq.~\eqref{mu_eq2} reduces to
\begin{align}
3(3m_{\rm eff}^2 -2\Lambda_g)\hat{\mu} 
+ \kappa_g^2 \tilde{\rho}_g
- K^2\kappa_f^2 \tilde{\rho}_f=0\,,
\end{align}
which fixes the solution of $\hat{\mu}$  by
\begin{align}
\hat{\mu}=-\frac{\kappa_g^2 \tilde{\rho}_g
- K^2\kappa_f^2 \tilde{\rho}_f}
{3(3m_{\rm eff}^2 -2\Lambda_g)}\,.
\end{align}
From this solution, 
the ansatz $|\hat{\mu}|\ll 1$ imposes the condition such that 
\begin{align}
3|3m_{\rm eff}^2 -2\Lambda_g| \gg 
|\kappa_g^2 \tilde{\rho}_g
- K^2\kappa_f^2 \tilde{\rho}_f| 
\label{Vainshtein}\,.
\end{align}

In the regime of small $\hat{\mu}$, one of the metric perturbation
is given by
\begin{align}
\frac{\hat{\Phi}'_g}{R} \approx 
\frac{1}{6}\Biggl[
\left(1+\frac{m_g^2}{3m_{\rm eff}^2-2\Lambda_g}\right)\kappa_g^2\tilde{\rho}_g 
-\frac{m_g^2}{3m_{\rm eff}^2-2\Lambda_g}K^2\kappa_f^2\tilde{\rho}_f\Biggl]
+\frac{1}{3}\Lambda_g\,,
\end{align}
which cannot be restored to the GR result  
even in the massless limit.

From Eq.~\eqref{Vainshtein}, the ansatz $|\hat{\mu}|\ll 1$  is no longer valid
for a high matter density.
In the high dense regime, the solution of Eq.~\eqref{mu_eq2} is given by 
the equation
\begin{align}
\mathcal{C}_{\rm matter}
\approx 0
\,,
\end{align}
from which the  typical value of $\hat\mu$  is the order of unity.
The  mass term
is negligible compared to the matter term 
in Eqs.~\eqref{flat_psig}-\eqref{flat_phig},
because $\kappa_g^2\tilde \rho_g \gg m_g^2$ and 
 $\kappa_f^2\tilde \rho_f \gg m_f^2$.
Therefore, the metric perturbations in the physical sector
(and also in the fiducial sector) are restored to the ones in GR.

However it is not clearly seen whether we can connect 
the two limiting cases, i.e., 
the low-dense bigravity region with $|\hat \mu|\ll1$ and 
the high-dense GR region with $|\hat \mu|\sim O(1)$.
For the screening, we must find a regular solution 
connecting both regions.
For simplicity, 
we assume that  the matter fields are 
localized in finite regions 
and only consider the outside of the matter distributions.
The spatial averages of the matter densities within the radius $R$ and $R_f$ 
are give by 
\begin{align}
\tilde{\rho}_g(R)
=\frac{3M_g}{4\pi R^3}
,\quad
\tilde{\rho}_f(R)=\frac{3\mathcal{M}_f}{4\pi R^3(1+\mu)^3}
\label{mean_density}
\,,
\end{align}
where the masses are defined by
\begin{align}
M_g=\int_0^{R_0} 4\pi\rho_g R^2 dR\,,~~
\mathcal{M}_f=\int_0^{R_{f,0}} 4\pi\rho_fR^2dR\,,
\end{align}
respectively.
Here $R_0$ and $R_{f,0}$ are the surface radii of matter distributions.
From Eqs.~(\ref{Vainshtein}) and (\ref{mean_density}), 
the Vainshtein radius, beyond which non-linearity of $\hat \mu$
becomes important,  is estimated as
\begin{align}
R_V\approx
\left(
\frac{2|GM_g-K^2\mathcal{GM}_f|}{3m_{\rm eff}^2-2\Lambda_g}
\right)^{1/3}\,.
\end{align}
We find that the Vainshtein radius depends on the cosmological constant
as well as the $f$-matter field.
Outside this radius, the smallness condition for $\hat{\mu}$ is valid.

Before introducing the $f$-matter, 
we summarize the case without the $f$-matter
as discussed in \cite{Vainshtein}.
Since we find the similar results in the case with a small mass object
of the $f$-matter, 
we explain the Vainshtein screening by our result.
 
\begin{figure*}[htbp]
\begin{center}
\includegraphics[width=15cm,angle=0,clip]{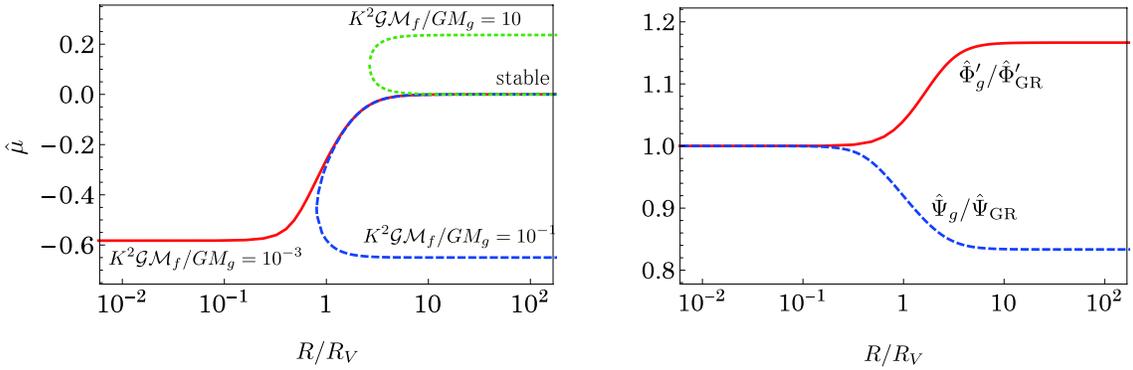}
\caption{The left figure shows 
asymptotically homothetic branches $(\hat{\mu}\rightarrow 0)$ 
of Eq.~\eqref{mu_eq2}.
We set $\beta_2=-3,\beta_3=3,m_g^2=m_f^2, \Lambda_g=0$,
 and $K^2\mathcal{GM}_f/GM_g=10^{-3}$ (red solid curve),
$K^2\mathcal{GM}_f/GM_g=10^{-1}$ (blue dashed curve) 
and $K^2\mathcal{GM}_f/GM_g=10$ (green dotted curve).
Except for $K^2\mathcal{GM}_f/GM_g=10^{-3}$,
the asymptotically homothetic branches do not 
extend  below the Vainshtein radius.
The right figure shows the existence of the Vainshtein screening
for $K^2\mathcal{GM}_f/GM_g=10^{-3}$, i.e., 
GR is recovered for $R<R_V$, where 
we define  $\hat{\Phi}'_{\rm GR}=\hat{\Phi}'_g|_{m_g=0}$
and $\hat{\Psi}_{\rm GR}=\hat{\Psi}_g|_{m_g=0}$.
}
\label{flat_Vainshtein_fig1}
\end{center}
\end{figure*}

\begin{figure*}[htbp]
\begin{center}
\includegraphics[width=15cm,angle=0,clip]{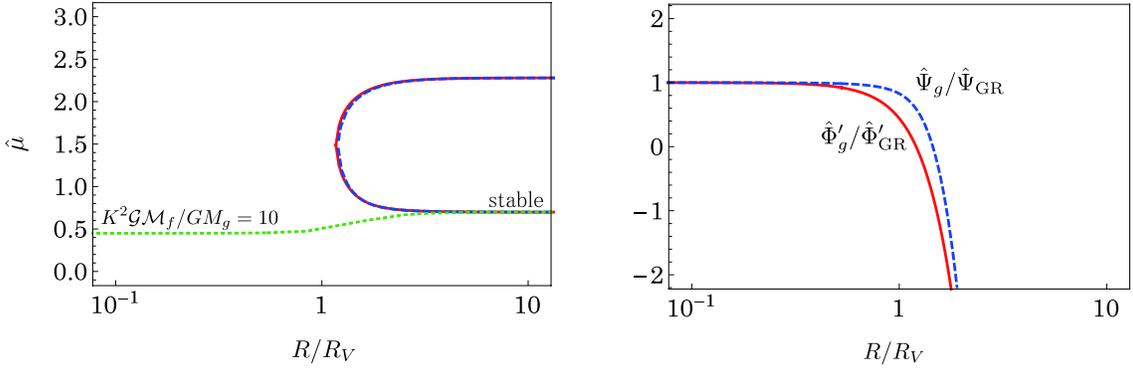}
\caption{The same as Fig.~\ref{flat_Vainshtein_fig1}
for non-homothetic branches.
The asymptotically stable non-homothetic branch
is characterized by
$\hat{\mu} \rightarrow 0.70$ as $R/R_V\rightarrow \infty$.
The deviation of bigravity from GR diverges as  
$\hat{\Psi}'_g/\hat{\Psi}'_{\rm GR}
\propto R^3$ beyond the Vainshtein radius
and similar to the $\hat{\Phi}_g/\hat{\Phi}_{\rm GR}$.
}
\label{flat_Vainshtein_fig2}
\end{center}
\end{figure*}
\end{widetext}

\begin{figure}[htbp]
\begin{center}
\includegraphics[width=7.5cm,angle=0,clip]{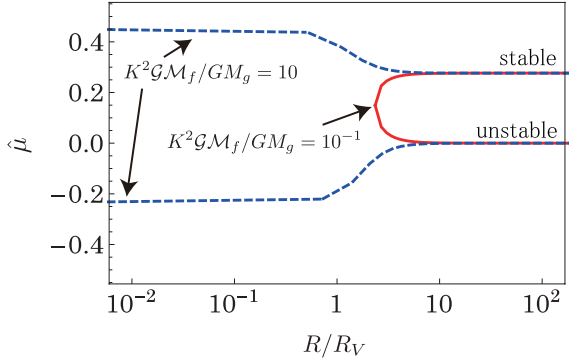}
\caption{The roots of Eq.~\eqref{mu_eq2}.
We set $\beta_2=-3,\beta_3=3,m_g^2=m_f^2$ and $\Lambda_g=10m_g^2$.
An asymptotically stable branch are characterized by
$\hat{\mu}\rightarrow \hat\mu_\infty =0.28$ as $R/R_V\rightarrow \infty$.
The gravitational property is almost same everywhere
as discussed in Sec.~\ref{sec_GR_phase}.
}
\label{dS_Vainshtein_fig}
\end{center}
\end{figure}

In Fig.~\ref{flat_Vainshtein_fig1}, we present our results 
in the case of $K^2 {\cal G}{\cal M}_f=10^{-3}
GM_g$.
The red curves in the figure show the value of $\hat \mu$ (left) 
and the ratio of the gravitational force to the one in GR (right).
From these figures, we find that GR is recovered within the 
Vainshtein radius $R_V$, but the bigravity phase appears in the 
region of $R>R_V$. 
This is just the Vainshtein screening.
The sufficiently small amount of the $f$-matter
does not spoil the Vainshtein mechanism.

We then discuss whether there exists a regular solution which
shows the Vainshtein screening if we have 
a large amount of the $f$-matter
 and/or  a cosmological constant.

First, we consider asymptotically homothetic solutions
without a cosmological constant.
As shown in Fig. \ref{flat_Vainshtein_fig1},
if the amount of the  $f$-matter field gets large,
the $\hat \mu$-curve cannot reach below the Vainshtein radius.
The derivative of $\hat{\mu}$ with respect to $R$ diverges around the Vainshtein radius,
and then  the assumption \eqref{assumption_flat_V} is broken there.
This hints a sharp transition from the linear to nonlinear regimes, and we must not neglect the gradient terms of $\hat\mu$
such as $R\hat{\mu}'\hat{\Phi}_g$ 
even if the metric perturbations ($|\hat{\Phi}_g|, |\hat{\Psi}_g|, |\hat{\Phi}_f|,$ and $
 |\hat{\Psi}_f|$) themselves are small.

There exists another solution which does not approach a homothetic solution 
as $R/R_V\gg 1$
 below the Compton wavelength of the massive graviton.
We call it a non-homothetic branch.
As shown in Fig.~\ref{flat_Vainshtein_fig2}, 
for $K^2\mathcal{GM}_f/GM_g=10$, we find a regular solution
which approaches a stable non-homothetic branch for $R\gg R_V$
(see Appendix \ref{app_late_stability} for the stability analysis).
In this branch, GR is recovered inside the Vainshtein radius, 
but the outside is quite different. 
The metric perturbations 
scale as $|\hat{\Psi}_g|, |\hat{\Phi}_g|\propto m_g^2R^2$ asymptotically.
 Note that,
while the spacetime is given by an inhomogeneous spacetime 
in $R \ll m_{\rm eff}^{-1}$,
we find the spacetime would approach  an anti-de Sitter spacetime
beyond the Compton wavelength of the massive graviton
by solving the static spacetime without the ansatzes \eqref{weak_gravity1} and \eqref{weak_gravity2}
\cite{preliminary}.
For a smaller amount of the $f$-matter, however, 
there is no such a regular solution.

Finally, we consider solutions with a  cosmological constant.
The behaviour of $\hat{\mu}$ is similar to the asymptotically flat case 
when $3m_{\rm eff}^2-2\Lambda_g\gtrsim 0$. 
However, if $3m_{\rm eff}^2-2\Lambda_g\lesssim 0$,
the behaviour of $\hat{\mu}$ changes qualitatively.
As shown in Fig.~\ref{dS_Vainshtein_fig},
when the amount of the $g$-matter is larger than that of the $f$-matter, 
the derivative $\hat{\mu}$ diverges around the Vainshtein radius,
indicating a sharp transition and the breakdown of our assumption of small gradients.
On the other hand, we can obtain regular solutions when
the amount of the $f$-matter is larger than that of the $g$-matter.
Note that while the  $\hat{\mu}_\infty=0$\,-\,branch is unstable,
the asymptotically $\hat{\mu}_\infty= 0.280$\,-\,branch is stable 
(see Appendix \ref{app_late_stability}
for the stability analysis).
For the stable branch, although GR is recovered within the Vainshtein radius, 
the outside is a stable non-homothetic spacetime. 
As discussed in Sec.~\ref{cosmology_with_nonlinear},
the gravitational behaviour is almost the same as in GR
when the cosmological constant is much larger than the graviton mass.

\subsection{Stability of a spherically-symmetric static spacetime}
\label{app_late_stability}

We now turn to analyzing the stability of the spherically-symmetric static solution obtained 
in the previous subsection
 in the region $R\gg R_V$, in which
the static solution is also obtained from the cosmological solution 
\eqref{mu_eq_without_matter}
in the limit $H\rightarrow 0$ or $H\rightarrow \Lambda_g/3$.
The matter effects can be ignored in $R\gg R_{V}$.
The perturbed metrics are given by
\begin{align}
ds_g^2&=-e^{2\hat{\Phi}_g+2\hat{\phi}_g}dT_g^2+e^{2\hat{\Psi}_g+2\hat{\psi}_g}dR_g^2+R_g^2d\Omega^2\,,\\
ds_f^2&=K^2\left[
-e^{2\hat{\Phi}_f+2\hat{\phi}_f}dT_f^2+ e^{2\hat{\Psi}_f+2\hat{\psi}_f}dR_f^2+R_f^2d\Omega^2\right]\,,
\end{align}
where
$(\Phi_g,\Psi_g,\Phi_f,\Psi_f)$ are determined by \eqref{flat_psig}-\eqref{flat_phif}
with $\hat{\mu}=\hat{\mu}_{\infty}$,
a root of $\mathcal{C}_{m^2}+\mathcal{C}_{\Lambda}=0$.
We set the gauge as
\begin{align}
T_g=T+\delta T\,, \quad
R_g=R+\delta R\,, \\
T_f=T\,,\quad
R_f=(1+\hat{\mu}_{\infty})R\,.
\end{align}

\begin{widetext}
We consider the quadratic action in terms of
$(\hat{\phi}_g,\hat{\psi}_g,\hat{\phi}_f,\hat{\psi}_f,\delta T, \delta R)$
with $\Lambda_gR^2\ll 1$ and $ m_{\rm eff} R\ll 1$,
namely the scales smaller than the cosmological horizon and the Compton wavelength of the graviton.
We find a constraint equation, which is solved  when we set 
\begin{align}
\delta T=-\frac{\partial \hat{\pi}}{\partial T}\,, \quad
\delta R=\frac{1}{1+\hat{\mu}_{\infty}}\frac{\partial \hat{\pi}}{\partial R}\,.
\end{align}
by introducing the St\"{u}ckelberg variable $\hat{\pi}$.
The perturbed metric variables are not dynamical. They are given explicitly by
\begin{align}
\hat{\psi}_g&=-\frac{m_g^2R}{2}(1+2\beta_2\hat{\mu}_{\infty}+\beta_3\hat{\mu}_{\infty}^2)\frac{\partial \hat{\pi}}{\partial R}
\,, \\
\frac{\partial \hat{\phi}_g}{\partial R}&=
-\frac{m_g^2}{2}\Biggl[
(-1+\beta_3\hat{\mu}_{\infty}^2)\frac{\partial \hat{\pi}}{\partial R}
+(1+2\beta_2\hat{\mu}_{\infty}+\beta_3\hat{\mu}_{\infty}^2)R
\frac{\partial^2 \hat{\pi}}{\partial T^2}
\Biggl]
\,,\nn
\hat{\psi}_f&=\frac{m_f^2R}{2(1+\hat{\mu}_{\infty})^2}(1+2\beta_2\hat{\mu}_{\infty}
+\beta_3\hat{\mu}_{\infty}^2)\frac{\partial \hat{\pi}}{\partial R}
\,,\\
\frac{\partial \hat{\phi}_f}{\partial R}&=
\frac{m_f^2}{2}\Biggl[
-\frac{1+2\hat{\mu}_{\infty}+(2\beta_2-\beta_3)\hat{\mu}_{\infty}^2}{(1+\hat{\mu}_{\infty})^2}
\frac{\partial \hat{\pi}}{\partial R}
+(1+2\beta_2\hat{\mu}_{\infty}+\beta_3\hat{\mu}_{\infty}^2)R\frac{\partial^2 
\hat{\pi}}{\partial T^2}
\Biggl]\,.
\end{align}
Then we obtain the quadratic action in terms of $\hat{\pi}$ as
\begin{align}
S_2= \frac{m_{\rm eff}^2}{2 \kappa_-^2} \int d\Omega
\int dT dR
R^2 \left[Z^{TT} (\partial_T \hat{\pi})^2
- Z^{RR}(\partial_R \hat{\pi})^2 \right]
\end{align}
where 
\begin{align}
Z^{TT}=& \frac{3}{2} m_g^2\left[ 
1+\left(6 \beta _2-3\right)\hat{\mu}_{\infty} 
+\left(6 \beta _2^2-6 \beta _2+4 \beta _3\right) \hat{\mu}_{\infty} ^2
-\frac{2}{3} \left(3 \beta _2^2-10 \beta _3 \beta_2+2 \beta _3\right) \hat{\mu}_{\infty} ^3
+\frac{5}{3} \beta _3^2 \hat{\mu}_{\infty} ^4
+\frac{1}{3} \beta _3^2 \hat{\mu}_{\infty} ^5
\right]
\nn
& + \frac{3}{2}\frac{m_f^2}{1+\hat{\mu}_{\infty}}
\biggl[
1+\left(6 \beta _2-2\right) \hat{\mu}_{\infty} 
+\left(6 \beta _2^2-2 \beta _2+4\beta _3-2\right) \hat{\mu}_{\infty} ^2
-\frac{2}{3} \left(2\beta _2^2+\left(3-10 \beta _3\right) \beta _2-2 \beta _3+1\right) \hat{\mu}_{\infty} ^3
\nn
&\qquad\qquad \qquad
+\frac{1}{3} \left(-2 \beta _2^2-2 \beta _2+\beta _3 \left(5 \beta _3+2\right)\right) \hat{\mu}_{\infty} ^4
\biggl]
\nn
& - \Lambda_g\left[
1
+\left(4 \beta _2-2\right) \hat{\mu}_{\infty}
+\left(-\beta _2+3 \beta _3-1\right) \hat{\mu}_{\infty} ^2
+\left(\beta _3-\beta _2\right) \hat{\mu}_{\infty} ^3
\right]\,, \\
Z^{RR}&=
\frac{3}{2}\frac{m_g^2}{1+\hat{\mu}_{\infty}}\biggl[
1
+\left(4 \beta _2-2\right) \hat{\mu}_{\infty} 
+\left(2 \beta _2^2-2 \beta _2+\frac{4 \beta_3}{3}\right) \hat{\mu}_{\infty} ^2
-\frac{5}{9} \beta _3^2 \hat{\mu}_{\infty} ^4
-\frac{2}{9} \beta _3^2 \hat{\mu}_{\infty} ^5
\biggl]
\nn
&
+\frac{3}{2}\frac{m_f^2}{(1+\hat{\mu}_{\infty})^3}
\biggl[
1
+4 \beta _2 \hat{\mu}_{\infty} 
+\frac{2}{3} \left(3 \beta _2^2+8 \beta_2+2 \beta _3-4\right) \hat{\mu}_{\infty} ^2
+\frac{4}{3} \left(3 \beta _2^2-\beta _2+2 \beta _3-1\right) \hat{\mu}_{\infty} ^3
\nn
&\qquad\qquad \qquad \quad
+\frac{1}{9} \left(-2 \beta _2^2+4 \left(5 \beta _3-1\right) \beta _2-5 \beta _3^2+2 \beta _3-2\right) \hat{\mu}_{\infty}^4
\biggl]
\nn
&
-\frac{\Lambda_g}{1+\hat{\mu}_{\infty}}
\left[
1
+\frac{4}{3} \left(2 \beta _2-1\right) \hat{\mu}_{\infty} 
+\left(-\frac{\beta _2}{3}+\beta _3-\frac{1}{3}\right) \hat{\mu}_{\infty} ^2
\right]\,.
\end{align}
\end{widetext}

The no-ghost and no-gradient instability conditions are
given by
\begin{align}
Z^{TT} > 0,\quad \frac{Z^{RR}}{Z^{TT}} > 0\,.
\end{align}
Note that for the linear branch $\hat{\mu}_{\infty}=0$, 
the coefficients become
\begin{align}
Z^{TT}& = Z^{RR}= \frac{1}{2} \left( 3m_{\rm eff}^2-2\Lambda_g \right)
\,,
\end{align}
which reproduces Higuchi bound.
We also note that
for $m_g^2,m_f^2 \ll \Lambda_g$,
the above expressions reproduce 
the results discussed in Sec.~\ref{sec_stability} with $w=-1$
by using \eqref{mu_pm} and the coordinate transformation.

For other branches of solutions for $\hat\mu$, since the coefficients are complicated
(where $\hat{\mu}_{\infty}$ is zero or the roots of the sixth-order algebraic equation
$\mathcal{C}_{m^2}+\mathcal{C}_{\Lambda}=0$),
we cannot obtain the explicit expressions for no-instability conditions in terms of 
$\beta_2, \beta_3, m_g, m_f,$ and $\Lambda_g$.
Hence, we check numerically 
whether the solutions in the nonlinear branches are stable or not
for some instructive sets of coupling constants.

\begin{figure}[htbp]
\begin{center}
\includegraphics[width=7cm,angle=0,clip]{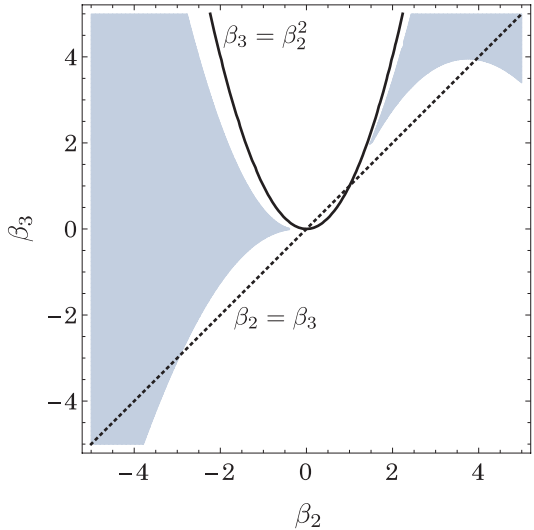}
\caption{
The parameter region of
the existence of a stable  asymptotically non-homothetic branch.
We set $m_g^2=m_f^2$ and $\Lambda_g=0$.
In the colored region, 
there is a stable branch
with $\hat{\mu}_{\infty}>-1$
other than $\hat{\mu}_{\infty}=0$.
The curves for $\beta_3=\beta_2^2$ (solid curve) and $\beta_3=\beta_2$ (dotted line) are also shown for the comparison with the stability condition \eqref{stable_condition} in the regime $H\gg m_{\rm eff}$, obtained in Subsec.~\ref{sec_stability}.
}
\label{no_instability_Min}
\end{center}
\end{figure}

As shown in Fig.~\ref{no_instability_Min},
the parameters can be classified into two types.
We show two typical examples in the parameter region \eqref{stable_condition}.

Case A: We choose
\begin{align*}
m_g=m_f\,, \quad \beta_2=-3\,, \quad \beta_3=3\,,
\end{align*}
in which we can also find a stable branch
other than $\hat{\mu}_{\infty}=0$ branch.
For instance, the nonlinear stable branch 
is given by
\begin{align*}
\hat{\mu}_{\infty} = 0.698872\,,
\end{align*}
with $\Lambda_g=0$.
For non-zero cosmological constant,
the stable branch is given by only
\begin{align*}
\hat{\mu}_{\infty} = 0.280938\,,
\end{align*}
with $\Lambda_g=10\, m_g^2$.

Case B: We choose
\begin{align*}
m_g=m_f\,, \quad \beta_2=-1\,, \quad \beta_3=\frac{1}{2}\,,
\end{align*}
in which only $\hat{\mu}_{\infty}=0$ is stable with $3m_{\rm eff}^2-2\Lambda_g\gtrsim 0$.
However, although 
$\hat{\mu}_{\infty}=0$ becomes unstable,
one stable branch appears when $3m_{\rm eff}^2-2\Lambda_g\lesssim 0$.
For instance, the nonlinear stable branch is given by
\begin{align*}
\hat{\mu}_{\infty}= 0.889668\,,
\end{align*}
with $\Lambda_g=10\, m_g^2$.


\end{document}